\documentclass[a4paper,12pt]{article}

\pdfoutput=1

\usepackage{amssymb,amsmath,bm}
\usepackage{a4wide}
\usepackage{color}
\usepackage{slashed}
\usepackage{cite}
\usepackage{graphicx}
\usepackage[latin1]{inputenc}
\usepackage{amsfonts}
\usepackage{lscape}
\usepackage{amsthm}
\usepackage{booktabs}
\usepackage{array}
\usepackage{rotating}
\usepackage{float}
\usepackage{multirow}
\usepackage{graphicx,rotating,amsmath,multirow,xcolor,colortbl,xcolor}
\usepackage{hyperref,pdflscape,slashed}
\definecolor{rosso}{cmyk}{0,1,1,0.4}
\definecolor{rossos}{cmyk}{0,1,1,0.55}
\definecolor{rossoc}{cmyk}{0,1,1,0.2}
\definecolor{blu}{cmyk}{1,1,0,0.3}
\definecolor{blus}{cmyk}{1,1,0,0.6}
\definecolor{bluc}{cmyk}{1,1,0,0.1}
\definecolor{verde}{cmyk}{0.92,0,0.59,0.25}
\definecolor{verdec}{cmyk}{0.92,0,0.59,0.15}
\definecolor{verdes}{cmyk}{0.92,0,0.59,0.4}

\newcommand{\MtEW}{ (177.0\pm2.6)\,{\rm GeV}}

\def\arXiv#1{{\color{rossos}\href{http://arxiv.org/abs/#1}{arXiv:#1}}} 
\hypersetup{colorlinks,bookmarksopen,bookmarksnumbered,
linkcolor=blus,pdfstartview=FitH,urlcolor=rossos,citecolor=verde}

\font\tenrsfs=rsfs10 at 12pt
\font\sevenrsfs=rsfs7
\font\fiversfs=rsfs5
\newfam\rsfsfam
\textfont\rsfsfam=\tenrsfs
\scriptfont\rsfsfam=\sevenrsfs
\scriptscriptfont\rsfsfam=\fiversfs
\def\mathscr#1{{\fam\rsfsfam\relax#1}}

\def\Lag{\mathscr{L}}

\definecolor{Gray}{gray}{0.95}
\newcommand{\bbox}[1]{\fcolorbox{gray}{Gray}{~$\displaystyle #1$~}}

\newcommand{\BR}{{\rm BR}}

\usepackage[small,bf]{caption}
\newcommand{\gev}{{\rm GeV}}
\newcommand{\mev}{{\rm MeV}}

\newcommand{\vcb}{|V_{cb}|}
\newcommand{\vtd}{|V_{td}|}
\newcommand{\vub}{|V_{ub}|}
\newcommand{\vts}{|V_{ts}|}
\newcommand{\vus}{|V_{us}|}

\newcommand{\eps}{\epsilon}

\def\kpn{K^+\rightarrow\pi^+\nu\bar\nu}
\def\klpn{K_{L}\rightarrow\pi^0\nu\bar\nu}

\newcommand{\be}{\begin{equation}}
\newcommand{\ee}{\end{equation}}
\newcommand{\bea}{\begin{eqnarray}}
\newcommand{\eea}{\end{eqnarray}}
\newcommand{\beq}{\begin{equation}}
\newcommand{\eeq}{\end{equation}}
\newcommand{\beqa}{\begin{eqnarray}}
\newcommand{\eeqa}{\end{eqnarray}}

\newcommand{\mtb}{\overline{m}_{t}}

\def\eq#1{eq.~(\ref{#1})}

\selectfont

\begin{document}

\centerline{CERN-PH-TH-2015-191\hfill IFUP-TH/2015}

\vspace{0.2truecm}

\begin{center}
\boldmath

{\textbf{\huge\color{rossos} Indirect determinations\\[4mm] of the top quark mass}} 
\unboldmath

\bigskip

\vspace{0.1truecm}

{\bf\large Gian F. Giudice$^a$, Paride Paradisi$^b$,
Alessandro Strumia$^{c}$}  \\[5mm]

{\it $^a$ CERN, Theory Division, Geneva, Switzerland}\\[0mm]
{\it $^b$ Dipartimento di Fisica dell'Universit{\`a} di Padova and INFN, Italy}\\[0mm]
{\it $^c$ Dipartimento di Fisica dell'Universit{\`a} di Pisa and INFN, Italy}\\[0mm]
\end{center}

\vspace{-1cm}

\begin{quote}\large
\noindent

We give a complete analysis of indirect determinations of the top quark mass in the Standard Model by introducing a systematic procedure to identify observables that receive quantum corrections enhanced by powers of $M_t$. 
We discuss how to use flavour physics as a tool to extract the top quark mass. 
Although present data give only a poor determination, we show 
how future theoretical and experimental progress in flavour physics can lead to an accuracy in $M_t$ well below 2 GeV. We revisit determinations of $M_t$ from electroweak data, 
showing how an improved measurement of the $W$ mass leads to an accuracy at the level of 1 GeV.

\end{quote}

\thispagestyle{empty}
\vfill

\vspace{-0.5cm}

\tableofcontents


\section{Introduction}

The top quark mass ($M_t$) is a key input parameter of the Standard Model (SM). Since the top quark is the heaviest particle in the SM, 
its Yukawa coupling $y_t$ is sizeable and plays a crucial role in determining the predictions of the theory at the quantum level.
A precise determination of $M_t$ is crucial for:
\begin{itemize}
\item  Stability of the electroweak vacuum. Assuming that no new physics modifies the short-distance behaviour of the SM, top-quark loops destabilise the Higgs potential creating a deeper minimum at large field value. The measured SM parameters lie so close to the critical condition for the formation of the large-field minimum that the instability scale can fluctuate from $10^{10}\, \gev$ to the Planck scale with a variation of $M_t$ of merely 2~GeV~\cite{higgsinst, higgsinst2}. Any such small change in $M_t$ can have a substantial effect in the evolution of the universe at the inflationary epoch~\cite{higgscosmo} and determine the viability of scenarios of Higgs inflation~\cite{higgsinf}. A more precise determination of $M_t$ will add important information to our knowledge of particle physics and cosmology.

\item  Supersymmetric predictions for the Higgs mass.
Within the Minimal Supersymmetric Standard Model, the soft-breaking scale that reproduces the observed Higgs mass
has a strong dependence on $M_t$.  For $\tan\beta=1$, the supersymmetry-breaking scale is large and roughly coincides with the SM stability scale discussed above.
For $\tan\beta=20$, maximal stop mixing and degenerate sparticles, 
precision computations~\cite{MhSUSY} find that the supersymmetry-breaking scale varies from
$1.7$ to $2.5$ TeV when $M_t$ is varied by one standard deviation around its present best-fit value.
\end{itemize}
The most precise quoted value of the top-quark pole mass comes from the
combination of  LHC and Tevatron measurements~\cite{ATLAS:2014wva}
\beq
(M_t)_{\rm pole} = 173.34 \pm 0.76~ \gev \, .
\label{topm}
\eeq
A theoretical concern about the extraction of $M_t$ from data is that the pole top mass is not a physical observable. 
This means that its experimental determination is done through the measurement of other physical observables (final-state invariant masses, kinematic distributions, total rates) that are especially sensitive to $M_t$. These measurements are compared to the results of theoretical calculations, which are expressed in terms of $M_t$ in a well-defined renormalisation scheme. In the context of hadron colliders, the extraction of $M_t$ suffers from a variety of effects linked to hadronization that are not fully accountable by perturbative QCD calculations, like bound-state effects of the $t\bar t$ pairs, parton showering, and other non-perturbative corrections (see~\cite{Juste:2013dsa} for a thorough discussion). In practice, the extraction of $M_t$ relies on modelling based on Monte-Carlo generators, and this is why~\cite{Agashe:2013hma} refers to $M_t$ in \eq{topm} as ``Monte-Carlo mass". Its relation to any short-distance definition of the top mass has an inherent ambiguity due to infrared non-perturbative effects, which probably amount to about 0.3~GeV. Much work is ongoing both on the experimental and the theoretical sides to control the size of the errors at this level.

\medskip

Alternative methods to extract $M_t$ have been proposed, with the aim of finding observables whose prediction is theoretically more robust. One interesting possibility is to identify observables that can be computed in QCD beyond the leading order in terms of the running top mass evaluated at a sufficiently high-energy scale, so that the perturbative expansion is completely reliable. The running top mass is then translated into the pole mass by means of a relation now known at four-loops in QCD~\cite{Marquard:2015qpa}. This programme has been applied to the total inclusive $t\bar t$ cross section~\cite{cross}, from which it was possible to extract the following values of the pole top mass: 
\beq \label{eq:Mtsigma}
(M_t)_{\sigma_{t\bar t}} = \left\{
\begin{array}{lcc}
172.9 \pm 2.6 ~\gev & \hbox{ATLAS} &\!\!\hbox{\cite{topatlas}}\\
176.7\pm 2.9 ~\gev & \hbox{CMS} &\!\! \hbox{\cite{topcms}}
\end{array}\right.   \ .
\eeq
Although the result is theoretically more transparent, the uncertainties in \eq{eq:Mtsigma} are still significantly larger than that  in \eq{topm}.

These considerations justify the search for alternative strategies to determine $M_t$, and this will be the subject of our paper.
Given that the top is the only quark associated to a sizeable Yukawa coupling, loop effects in the SM are potentially very sensitive to $M_t$.
Our goal is to identify all processes that receive quantum corrections enhanced by powers of $M_t$ (in the limit $M_t\gg M_W$) and infer $M_t$ from their measurements. 

\medskip

With the experimental confirmation that the Cabibbo-Kobayashi-Maskawa (CKM) matrix gives an overall successful explanation of the transitions among different quark generations, 
the main interest of flavour physics has turned towards the search for new effects beyond the SM. Indeed, flavour physics provides a unique tool to explore indirectly new physics, in a way often complementary 
to high-energy probes at colliders. However, in this paper we want to argue that new developments are guiding us towards a novel use of flavour physics data. On the experimental side, the lack of anomalous signals from the LHC suggests that new physics may lie at energy scales much higher than previously expected. On the theoretical side, present or upcoming improved calculations of flavour processes in the 
SM are opening new frontiers in precision measurements. In light of these developments, in this paper {\it we propose to use the comparison between experimental data and theoretical predictions of flavour processes as a way to extract the top quark mass}, under the assumption that the SM is valid up to very short distance scales.\footnote{For an earlier attempt to determine the top mass from $B$--$\bar B$ and the rare kaon decays $\kpn$, $\klpn$, see ref.~\cite{Buras:1992uf}.}

\smallskip

Our strategy is not new: the history of predicting quark masses from loop-induced flavour processes is glorious, with some of these predictions made even before the actual discovery of the corresponding particle. This is the case of the charm-quark mass, whose value was inferred from theoretical considerations on $K$--$\bar K$ mixing~\cite{Gaillard:1974hs} or of the top-quark mass, extracted using 
$B$--$\bar B$ data~\cite{Ellis:1977uk}.
The use of {flavour} data for an indirect determination of $M_t$ is fairly robust from the theoretical point of view, since it relies on controllable SM calculations, in which non-perturbative effects are restricted 
to a few well-known hadronic parameters, now under careful scrutiny by lattice calculations. 
In this paper, we describe the status of the extraction of the top mass from the fit of flavour data, finding 
$(M_t)_{\rm flavour} = (173.4 \pm 7.8)~ \gev$.
The uncertainty of this extraction is too large to be competitive with the direct measurements.
However,
taking into account foreseeable progress in perturbative and lattice calculations, on one side, and experimental measurements, on the other 
side, our projection for the future is that the error can be brought to about $1.7\,\gev$.

In our analysis we use the pole top mass $M_t$ as the physical quantity extracted from the fits, deriving it, whenever is needed, from the running $\overline{\rm MS}$ top mass $\mtb(m_t)$ through the $\mathcal{O}(\alpha_s^4)$ perturbative expression given in section~\ref{flobs}. This choice is dictated mostly by our desire to make the results more transparent and to adopt the same variable currently used by experimentalists. However, given that the pole mass, unlike the running $\overline{\rm MS}$, suffers from an $\mathcal{O}(\Lambda_{\rm QCD})$ inherent ambiguity, it may become more appropriate in the future, when higher accuracy is reached, to modify this choice, abandoning $M_t$ in favour of $\mtb(m_t)$.

An important byproduct of our analysis is that the top-mass extraction can be regarded as a well-defined motivation for improved experimental 
measurements and theoretical calculations in flavour physics. While the exploration for new-physics effects remains the most exciting part of 
the flavour physics programme, the extraction of $M_t$ defines a clear and concrete benchmark that can be used to determine the goals that 
experimental and theoretical improvements should aim for. 

\medskip

With the aim of an indirect determination of the top mass, in this paper we also reconsider global fits of electroweak observables, finding $(M_t)_{\rm EW} = \MtEW$, 
in good agreement with  previous studies~\cite{Baak:2014ora,Silvestrini}. We find that the determination of $M_t$ is dominated by the measurement of $M_W$.
A reduction of the error in the measurement of $M_W$ to about 8~MeV, as foreseeable at the LHC~\cite{Baak:2014ora}, can bring down the uncertainty on $M_t$ to $1.2\, \gev$.

Most of our considerations would be superseded by a futuristic $e^+e^-$ collider operating at the $t\bar t$ threshold. Such a collider would allow for an unprecedented 
determination of the top mass. Scans of the $t\bar t$ pair production would reach a statistical accuracy on the mass measurement of about 20--30 $\mev$~\cite{scans}. 
Recent N$^3$LO calculations can relate such measurements to a well-defined $M_t$, with a theoretical uncertainty below about $50\,\mev$~\cite{Beneke:2015kwa}.

\medskip

Our paper is organised as follows. In section~\ref{larget} we present a systematic procedure to identify observables sensitive, at the quantum level, to powers of the top mass.
We discuss present and future top mass determinations from flavour data
in section~\ref{flobs} and \ref{Mt_future}, respectively, and from electroweak precision data in section~\ref{secEW}. 
Conclusions are given in section~\ref{end}.

\section{$M_t$ dependence of observables in the heavy-top limit}
\label{larget}

The large top Yukawa coupling offers the possibility of reconstructing $M_t$ from SM quantum effects. In order to identify the physical observables that are most sensitive to the top mass at the one-loop level, we develop here a systematic procedure to extract the leading $M_t$ dependence predicted by the SM. We work in the heavy-top limit~\cite{heavyMt}, in which the masses of the $W$ and $Z$ bosons are neglected with respect to $M_t$. This is achieved by considering a gauge-less theory with massive quarks, the Higgs boson $h$, and 3 Goldstone bosons $\vec{\chi}$ (related by the equivalence theorem~\cite{equivth} to the longitudinal components of the $W$ and $Z$), where the only quark interaction is 
\beq
\mathscr{L} = y_t\, \bar{t}_R \, H^T \begin{pmatrix} V_{t i}\, d_{iL} \\ -t_L \end{pmatrix} +{\rm h.c.}
\label{lintt}
\eeq
Here $y_t$ is the top Yukawa coupling, $V$ is the CKM matrix, and we are working in a basis in which both quark mass matrices are simultaneously diagonal. The Higgs doublet $H$ is given by
\beq
H=\frac{1}{\sqrt{2}}e^{\frac{i\vec{\sigma} \cdot\vec{\chi}}{v}} \begin{pmatrix} 0 \\ v+h \end{pmatrix} \, ,
\eeq
where $v=246\, \gev$ is the symmetry breaking scale. We can explicitly write \eq{lintt} as
\bea
\Lag = &-&\frac{y_t}{\sqrt{2}}
\left( \cos{|\vec{\chi}|}/{v}\right) 
(v+h) \, \bar t t \nonumber \\
&+&y_t  \left( \frac{ \sin{|\vec{\chi}|}/{v}}{{|\vec{\chi}|}/{v}}\right)
\left( 1+\frac{h}{v}\right)\left[ \frac{i}{\sqrt{2}} \chi^0 \bar t \gamma_5 t +\left( \chi^+ \bar t_R V_{t i}\, d_{iL} + {\rm h.c.}\right) \right] \, ,
\label{picco}
\eea
where $\chi^0$ and $\chi^\pm$ are the neutral and charged Goldstones, and $|\vec{\chi}|^2={\chi^{0}}^2 +2\chi^+\chi^-$.
The next step is to integrate out the top quark using the interactions in \eq{picco}. The top-less effective theory will contain a set of effective operators 
whose coefficients readily describe the leading top-mass dependence in the large $M_t$ limit.

\begin{figure}[t]
\centering
\includegraphics[width=0.9\textwidth]{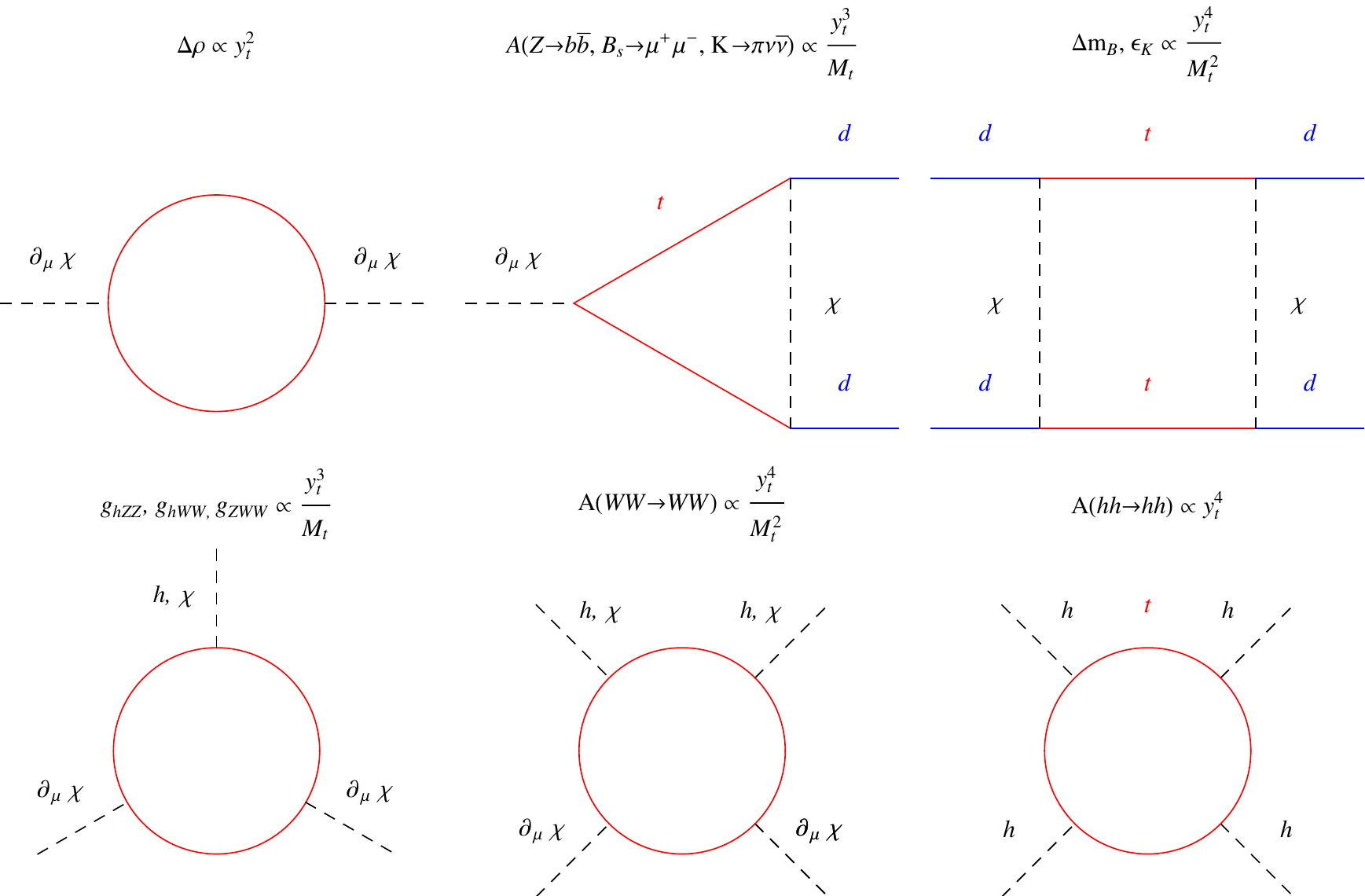}
\caption{\em Feynman diagrams illustrating the effective operators generated by integrating out the top quark. Also shown is the power counting estimate of their sensitivity to the top mass. 
Dashed lines denote the Higgs boson ($h$) or the Goldstones ($\chi$); solid lines denote the quarks.
\label{FeynMt}}
\end{figure}

\boldmath
\subsection*{$\Delta \rho$}
\unboldmath
At the level of dimension-4 operators, the first diagram in fig.~\ref{FeynMt} leads to a wave-function renormalisation of the Goldstone fields 
that violates the custodial SU(2) symmetry under which $\vec \chi$ transforms as a triplet. Simple power counting shows that this correction is $\mathcal{O}(y_t^2/16\pi^2)$, so we expect a quadratic sensitivity to $M_t$. Indeed, explicit calculation of the diagram in fig.~\ref{FeynMt} (together with a one-loop diagram obtained from the ${\vec \chi}\,{}^2\bar t t$ vertex, needed to cancel contributions at zero external momentum) reproduces the well-known result for the correction to the parameter $\rho  \equiv{M^2_W}/{\cos\theta_{\rm W}^2 M^2_Z}= 1+\Delta \rho$
\beq
\Delta \rho = \frac{3y_t^2}{32\pi^2}= \frac{3G_{\rm F} M^2_t}{8\sqrt{2}\pi^2} \,.
\eeq

\boldmath
\subsection*{$Z\to b\bar b$, $K\to\pi\nu\bar\nu$ and $B_s \to \ell^+\ell^-$}
\unboldmath
At the level of dimension-5 operators, the second diagram in fig.~\ref{FeynMt} leads to an effective coupling 
$ \bar d_L \gamma^\mu d_L(\partial_\mu  \chi^0)$ 
between a left-handed down current and the derivative 
of the neutral Goldstone $\chi^0$, which affects the $Z$ couplings.
By power counting we estimate the coefficient of the dimension-5 operator to be {of order} $|V_{td}|^2 y_t^3/(16\pi^2M_t)$, 
which corresponds again to a quadratic sensitivity on $M_t$. Explicit calculation of the diagram in fig.~\ref{FeynMt} gives a 
correction to the $Z\bar d_i d_j$ vertex
\beq
\Delta g_L^{ij} = \frac{V^*_{ti}V_{tj}\, y_t^2}{32\pi^2}\, .
\label{deltagg}
\eeq
The coupling $g_L$ is defined from
\beq   
\frac{g}{\cos\theta_{\rm W}}\, \bar d_i \left(\, g_L^{ij} P_L + g_R^{ij} P_R \,\right)\slashed{Z} d_j\,,
\label{Zff}
\eeq
where in the SM at tree level
\beq 
g_L^{ij} = \left( - \frac12 + \frac{\sin^2\theta_{\rm W}}{3} \right) \delta_{ij} \,,
\qquad\qquad
g_R^{ij} = \frac{\sin^2\theta_{\rm W}}{3} \delta_{ij}\,.
\eeq

The vertex correction in \eq{deltagg} gives a quadratic sensitivity to $M_t$ in the $Z \to b \bar b$ decay width
\beq
\Gamma(Z \to b \bar b) =  \frac{\rho \, G_{\rm F} M^3_Z}{\pi \sqrt{2}} 
\left[   (g_{L}^{bb}+ \Delta g_{L}^{bb})^2 +   (g_{R}^{bb})^2 \right] ,
\eeq
(here given for simplicity in the limit of vanishing bottom mass and neglecting QCD corrections)
and in the contribution to
the effective Hamiltonians describing $K\to\pi\nu\bar\nu$ and $B_s\to \ell^+\ell^-$ 
\begin{align}
\mathcal{H}^{\rm eff}_{K\to\pi\nu\bar\nu} &= \frac{\Delta g_L^{sd}}{2v^2}\,
(\bar{s}_L \gamma^\mu d_L )(\bar{\nu}^\ell_L \gamma_\mu \nu^\ell_L ) + \hbox{h.c.} \,,
\label{H_kpnn}\\
\mathcal{H}^{\rm eff}_{B_s \to \ell^+\ell^-} &= -\frac{\Delta g_L^{bs}}{2v^2}\, (\bar{b}_L \gamma^\mu s_L )(\bar\ell_L \gamma_\mu \ell_L ) + \hbox{h.c.}
\label{H_bsll}
\end{align}
The effects of eq.s~(\ref{H_kpnn})--(\ref{H_bsll}) in the corresponding branching ratios grow as $M_t^4$.
These results agree with the leading $M_t$ term of the known full one-loop calculation in the SM.

\boldmath
\subsection*{$\Delta m_{B_q}$ and $\epsilon_K$}
\unboldmath

The third diagram in  fig.~\ref{FeynMt} leads to a dimension-6 operator involving four $d_L$ fields. The estimate of the coefficient is $(V^*_{ti}V_{tj})^2y_t^4/(16\pi^2M_t^2)$, 
exhibiting quadratic sensitivity to the top mass. Computing the diagram in  fig.~\ref{FeynMt}, we find the $\Delta F = 2$ interaction
\begin{align}
&\mathcal{H}^{\rm eff}_{\Delta F = 2} =
\frac{y_t^2 (V^*_{ti}V_{tj})^2}{256\pi^2\, v^2}
(\bar{d}_{iL} \gamma^\mu d_{jL} )(\bar{d}_{iL} \gamma_\mu d_{jL} ) +\hbox{ h.c.}
\label{DeltaF2}
\end{align}
This gives a contribution to CP-conserving and CP-violating observables in meson--antimeson mixing with quadratic sensitivity on $M_t$, in the heavy-top limit. 
On the other hand, the charm-top one loop contribution to $\epsilon_K$ has no power sensitivity on $M_t$, in agreement with the full SM result.

\boldmath
\subsection*{Triple gauge boson vertices and $WW$ scattering}
\unboldmath

The diagrams in the bottom row of fig.~\ref{FeynMt} yield a variety of dimension-5 or dimension-6 operators involving $\chi$, $h$ and derivatives,
such as $h(\partial_\mu \chi)^2$, $\chi(\partial_\mu \chi)^2$, $h^2(\partial_\mu \chi)^2$, and $\chi^2(\partial_\mu \chi)^2$.
The usual power counting shows that they have a quadratic sensitivity on the top mass.\footnote{There is also an $\mathcal{O}(y_t^4/16\pi^2)$ correction to  $hh\to hh$ scattering and to the triple Higgs coupling. The sensitivity of the Higgs self-coupling to $y_t^4$ at the quantum level explains the importance of the top-mass measurement for vacuum stability considerations and for the calculation of the Higgs mass in supersymmetry.} 
These operators contribute to physical observables in triple gauge boson vertices and $WW$ scattering. Experimental sensitivity to these effects is too poor to allow for any significant determination of $M_t$. For this reason, we disregard these processes in our analysis, albeit their $M_t^2$ dependence.

\boldmath
\subsection*{$B\to X_s\gamma$}
\unboldmath
With the rules of the heavy-top effective theory, it is also easy to identify processes which have no power sensitivity on $M_t$. Such processes lead to poor determinations of $M_t$ because, in the large $M_t$ limit, 
one finds at best logarithmic dependences on the top mass.  One example is $B\to X_s\gamma$, for which the coefficient of the corresponding dimension-6 operator $m_b \bar s_L \sigma^{\mu\nu} b_R F_{\mu\nu}$ 
is estimated to be $eV_{tb}V^*_{ts}y_t^2/(16\pi^2 M_t^2)$. The lack of power sensitivity on $M_t$ is confirmed by the full result~\cite{Misiak:2015xwa,Czakon:2015exa} which, for $M_t$ in the vicinity of its physical value, gives
\beq 
\BR(B\to X_s\gamma) \propto 
\left(\frac{M_t}{173.34~\gev}\right)^{0.38}\, .
\eeq
For this reason, we will not include $B\to X_s\gamma$ in our analysis.

\boldmath
\subsection*{Higgs physics}
\unboldmath
The heavy-top effective theory also shows that, at present, Higgs physics is not a useful player in the game of extracting $M_t$.
The Higgs decays $h\to \gamma\gamma,\gamma Z$ can be induced by the operators $ h F_{\mu\nu}^2$, $h F_{\mu\nu} Z_{\mu\nu}$,
$h (\partial_\mu \chi^0) \partial_\nu F_{\mu\nu}$, whose coefficients are estimated to be $e^2y_t/(16\pi^2 M_t)$ (for the first two, which are dimension-5) and $ey_t^2/(16\pi^2 M_t^2)$ (for the third, 
which is dimension-6). This corresponds to the well-known result that the amplitudes for $h\to \gamma\gamma,\gamma Z$ quickly saturate in the large $M_t$ limit. Indeed, from the full SM result 
we find, for $M_t$ around its physical value,
\beq
\Gamma(h\to \gamma\gamma)\propto \left(\frac{M_t}{173.34~\gev}\right)^{0.037},\qquad
\Gamma(h\to Z\gamma)\propto \left(\frac{M_t}{173.34~\gev}\right)^{0.014}.
\eeq
For the same reason, also $h \leftrightarrow gg$ offers negligible sensitivity to variations of $M_t$ around its physical value. 

Another potential effect comes from the dimension-5 operator $h(\partial_\mu\chi)^2$, generated by the first Feynman diagram in the bottom row of fig.~\ref{FeynMt}, whose coefficient is $\mathcal{O}(y_t^3/16\pi^2M_t)$. An explicit evaluation of the diagram gives the following correction to the Higgs decay width into weak gauge bosons
\beq 
\frac{\Delta \Gamma(h\to WW^*,ZZ^*)}{\Gamma(h\to WW^*,ZZ^*)}= - \frac{5y_t^2}{32\pi^2} \, ,
\eeq
which agrees at the leading order in $y_t$ with the known SM result~\cite{Spira}.
Even a futuristic measurement of the branching ratio at 1\% could not determine $M_t$ with an error better than 50~GeV. 
The decays $h\to ZZ,WW$, in spite of their quadratic sensitivity on the top mass, in practice give no probe of $M_t$ because they are dominated by tree-level effects.

The process in which the Higgs is radiated off a $t\bar t$ pair offers a direct measurement of the top Yukawa coupling. However, the predicted precision in the determination of the ratio between the Higgs couplings to top and gluon is in the range 13--17\% for the LHC with 300~fb$^{-1}$ and 6--8\% at HL-LHC with 3000~fb$^{-1}$~\cite{hllhc}. This will never become competitive with other methods for extracting $M_t$ available in the future. More interesting is the case of a hadron collider at 100~TeV, where studies of the ratio $t\bar th/t \bar tZ$ 
could lead to a determination of the top Yukawa with one-percent accuracy.

\bigskip

We conclude this section by remarking how our analysis based on the heavy-top effective theory, after integrating out the top with interactions given in \eq{picco}, was useful to identify the observables most sensitive to $M_t$. However, for deriving quantitative results on $M_t$ and obtain reliable determinations, we have to turn to the full SM expressions of the relevant observables.

\section{Extracting $M_t$ from flavour data}
\label{flobs}
\begin{figure}[t]
\centering
\includegraphics[scale=0.5]{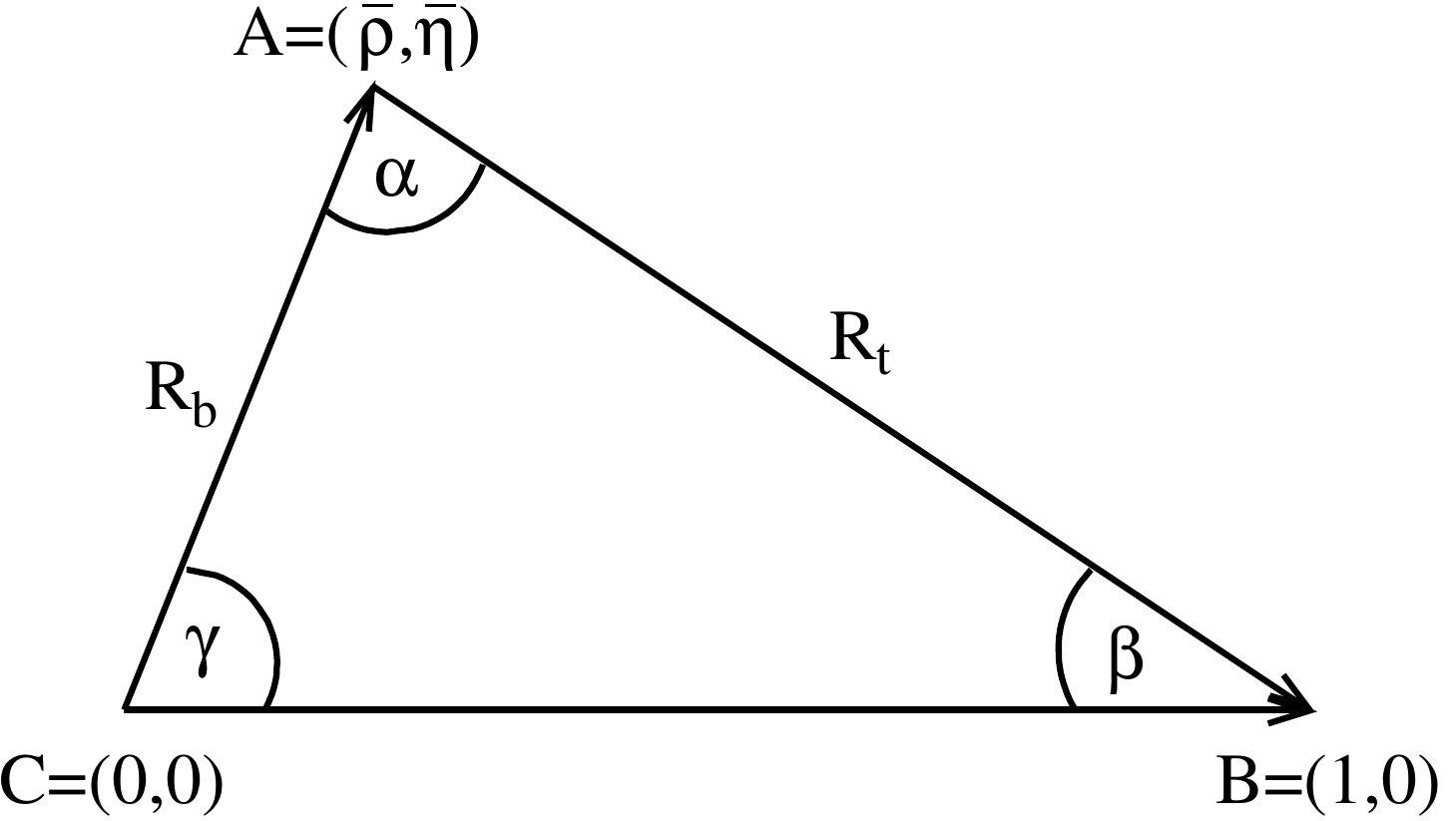}
\caption{\em The unitarity triangle.}
\label{fig:UT_triangle}
\end{figure}

We start by reviewing the basic relations among CKM matrix elements needed for our study.
Defining the four parameters $\lambda$, $A$, $\rho$, $\eta$ as
\beq
\lambda \equiv \frac{\vus}{\sqrt{|V_{ud}|^2+\vus^2}}\, , ~~~A\equiv \frac{\vcb}{\lambda \vus} \, , ~~~\varrho -i \eta \equiv \frac{V_{ub}}{A\lambda^3}\, ,
\eeq
the CKM matrix in the Wolfenstein parametrisation~\cite{Wolfenstein:1983yz} becomes
{\small \bea
\label{vmatr}
& ~~~~V \!\equiv\!
\begin{pmatrix}
V_{ud} & V_{us} & V_{ub}\\ 
V_{cd} & V_{cs} & V_{cb}\\ 
V_{td} & V_{ts} & V_{tb}\\ 
\end{pmatrix}
= &\\ &
\begin{pmatrix}
1 \!-\!{\lambda^2\over 2} \!-\!{\lambda^4\over 8} +\mathcal{O}(\lambda^6) & \lambda +\mathcal{O}(\lambda^7) & A\lambda^3(\varrho \!-\! i\eta)\\
-\lambda \!+\! A^2\lambda^5(\frac12 \!-\! \varrho \!-\! i\eta) + \mathcal{O}(\lambda^7)& 1\!-\!{\lambda^2\over 2} \!-\! {\lambda^4\over 2} (\frac14 \!+\! A^2)+\mathcal{O}(\lambda^6) & A\lambda^2 +\mathcal{O}(\lambda^8) \\
A\lambda^3(1 \!-\! \varrho \!-\! i \eta) +\frac{A\lambda^5}{2}( \varrho \!+\! i \eta) +\mathcal{O}(\lambda^7)& -A\lambda^2 \!+\! A\lambda^4  (\frac 12-\varrho \!-\! i\eta) +\mathcal{O}(\lambda^6)& 1 \!-\! \frac{ A^2 \lambda^4}{2}+\mathcal{O}(\lambda^6)  
\end{pmatrix} 
\,,
&
\nonumber
\eea}
in agreement with previous results~\cite{Buras:1994ec}.
Unitarity yields the condition
\begin{equation}
\label{UT_relation}
\frac{V_{ud}^{}V_{ub}^*}{V_{cd}^{}V_{cb}^*}   + \frac{V_{td}^{}V_{tb}^*}{V_{cd}^{}V_{cb}^*} +1=0\,,
\end{equation}
which can be represented as a triangle in the complex plane, see fig.~\ref{fig:UT_triangle}.
The vertex $A$ of the triangle is given by
\beq
\bar \varrho + i \bar \eta \equiv - \frac{V_{ud}^{}V_{ub}^*}{V_{cd}^{}V_{cb}^*} = (\varrho + i \eta) 
\left\{ 1-\frac{\lambda^2}{2} +\lambda^4\left[ A^2\left(\frac 12 -\varrho - i \eta\right) -\frac 18\right] +\mathcal{O}(\lambda^6)\right\} \, ,
\label{barbar}
\eeq
while the lengths of the sides $CA$ and $BA$, denoted by $R_b$ and $R_t$, respectively, are given by
\begin{eqnarray}
\label{R_b}
R_b &\equiv& \frac{| V_{ud}^{}V^*_{ub}|}{| V_{cd}^{}V^*_{cb}|}
= \sqrt{\bar\varrho^2 +\bar\eta^2}
=
 \frac{ \vub}{\lambda \vcb} \left[ 1-\frac{\lambda^2}{2} +\mathcal{O}(\lambda^4)\right] \,,\\
\label{R_t}
R_t &\equiv& \frac{| V_{td}^{}V^*_{tb}|}{| V_{cd}^{}V^*_{cb}|} =
 \sqrt{(1-\bar\varrho)^2 +\bar\eta^2}
= \frac{\vtd}{\lambda \vcb} \left[ 1 +\mathcal{O}(\lambda^4)\right] \,. 
\end{eqnarray}
The angles $\beta$ and $\gamma$ of the triangle are given by the expressions 
\begin{equation}
\label{R_tb_bis}
R_b=\frac{\sin(\beta)}{\sin(\gamma+\beta)}\,,\qquad
R_t= \frac{\sin(\gamma)}{\sin(\gamma+\beta)}\,,
\end{equation}
which allow us to write the coordinates of the vertex $A$ as 
\be
\label{etab_rhob}
\bar\varrho = R_b \cos\gamma = 1-R_t \cos\beta
\,,\qquad
\bar\eta = R_b \sin\gamma = R_t \sin\beta \,,
\ee
or, equivalently,
\beq
\frac{1}{\bar \varrho}=1+\frac{\tan\gamma}{\tan\beta}\, , \qquad
\frac{1}{\bar \eta}=\frac{1}{\tan\beta}+\frac{1}{\tan\gamma}\, .
\label{rhoeta}
\eeq

When searching for new physics, it is customary to determine the four independent CKM parameters from tree-level observables, 
which are presumed to be well described by the SM, and then use this determination to predict loop processes, which are expected 
to hide new effects beyond the SM.

In this paper, we take a different perspective: we are assuming the SM to be exactly valid and 
we are interested in extracting $M_t$ from flavour processes.  We then fix the four CKM parameters from the most precise measurements that 
do not depend on $M_t$, even if they arise at loop level\footnote{An alternative to our choice of CKM input variables could be to take $\vub$ instead of $\gamma$, since present relative errors of these two quantities are comparable, see table~\ref{table:inputs}. We prefer the choice in \eq{ckm_parameters_ours} for two reasons. First, $\vub$ induces larger uncertainties in the CKM combinations relevant to our analysis. Second, $\gamma$ is expected to be determined more precisely than $\vub$ in the future, see table~\ref{table:inputs}. Thus, we treat $\vub$ as a derived quantity, obtained from  $\vub= \vus\vcb {\sin\beta}/{\sin(\gamma+\beta)}[1+\lambda^2/2 +\mathcal{O}(\lambda^4)]$.}:
\begin{align}
\label{ckm_parameters_ours}
&\vus,\qquad \vcb, \qquad \gamma, \qquad \beta\,.
\end{align}
The parameters $\lambda$ and $A$ are related to $\vus$ and $\vcb$ in the usual way, while the expressions of $\varrho$ and $\eta$ in terms 
of $\gamma$ and $\beta$ are given in eq.s~(\ref{barbar}) and (\ref{rhoeta}). With this prescription, any element of the CKM matrix in \eq{vmatr} 
can be expressed in terms of the parameters in \eq{ckm_parameters_ours}. In particular, for our analysis we will need the following combinations
\bea
\label{Vtd}
|V_{td}V_{tb}^*| &=& \vus \vcb \,\frac{\sin \gamma}{\sin(\gamma +\beta )}\left[ 1+\mathcal{O}(\lambda^4)\right]\,,
\\
|V_{ts}V_{tb}^*| &=& \vcb \left[ 1 -\frac{\lambda^2\, \sin(\gamma -\beta)}{2\, \sin(\gamma + \beta )} +\mathcal{O}(\lambda^4)\right]\,,
\label{pirulon}
\\
\label{pirulin}
{\rm Re}\lambda_t &=&-{\vcb^2 \vus} \,\frac{\sin\gamma\, \cos \beta}{\sin (\gamma +\beta )}\left[ 1+\lambda^2\left( \frac 12 -\frac{\sin\gamma}{\cos\beta \sin(\gamma +\beta)}\right) +\mathcal{O}(\lambda^4)\right]\,,\\
{\rm Re}\lambda_c &=& -\vus \left[ 1-\frac{\lambda^2}{2}+\mathcal{O}(\lambda^4)\right]\,, \\
{\rm Im}\lambda_t &=& -{\rm Im}\lambda_c ={\vcb^2 \vus} \,\frac{\sin\gamma\, \sin \beta}{\sin (\gamma +\beta )} \left[ 1+\frac{\lambda^2}{2}+\mathcal{O}(\lambda^4)\right]\,,
\label{eq:Im_lambda_t}
\eea
where $\lambda_i=V_{id}^{}V_{is}^*$ (with $i=c,t$). 

\bigskip

Since the SM predictions for flavour observables are often expressed in terms of the running $\overline{\rm MS}$ top quark 
mass $\mtb(m_t)$, it is useful to give here the relation between the pole top mass $M_t$ and $\mtb(m_t)$.
Accounting for QCD corrections only\footnote{This is appropriate for flavour effects where higher-order electroweak corrections have not yet been computed.
When electroweak corrections have been computed, the result is expressed in terms of $M_t$, such that no conversion in necessary.}
we find~\cite{Marquard:2015qpa}
\begin{eqnarray}
\frac{M_t}{\mtb(m_t)} =  1 + 0.4244 \,{\alpha_s} + 0.8345 \,{\alpha_s^2} + 2.375 \,{\alpha_s^3} + (8.49 \pm 0.25) \,{\alpha_s^4}  = 1.060302(35)\,,
\label{mt}
\end{eqnarray}
where ${\alpha_s}\equiv \alpha_s^{(6)}(m_t) = 0.1088$. As remarked in the introduction, our choice to express results in terms of $M_t$ follows from standard practice, but has the disadvantage of using a quantity that it is affected by non-perturbative 
uncertainties of order $\Lambda_{\rm QCD}$.

We can now proceed the to discuss the extraction of the top mass from various flavour processes sensitive to $M_t$.

\def\arraystretch{0.95}
\begin{table}
\begin{center}
\begin{tabular}{c|c@{\hspace{-0.5em}}cc@{\hspace{-1.5em}}cc@{\hspace{-1.5em}}c}

Observable                 & Now (2015) &  & Error 2020 & & Error 2025 \\
\hline
$|G_{\rm F}|\!\times\! 10^{-5} (\gev^{-2})$   &  $1.16637(1)$ &  \cite{Agashe:2014kda} & -- & & -- &   \\
$M_W (\gev)$   &  $80.385(15)$ & \cite{Agashe:2014kda} & 8 & \cite{Baak:2014ora} & 5 &\cite{Baer:2013cma}   \\
$M_Z (\gev)$   &  {$91.1876(21)$} & \cite{Agashe:2014kda} & -- &  &  & --   \\
$\sin^2\theta_{\rm W}$   &  $0.23116(13)$ &   \cite{Agashe:2014kda} & 13 & \cite{Baak:2014ora} & 1.3 & \cite{Baak:2014ora}  \\
$\alpha_{\rm em}^{-1}(M_Z)$ &  128.952(13) &\cite{Agashe:2014kda} &--&&--\\
$\alpha_s(M_Z)$   &  $0.1184(7)$ &   \cite{Agashe:2014kda} & 7 & \cite{Baak:2014ora} & 7 & \cite{Baak:2014ora}  \\
$m_c(m_c) (\gev)$   &  $1.279(13)$ &   \cite{Chetyrkin:2009fv} & -- &  & --   \\
$m_K (\mev)$   &  $497.614(24)$ &   \cite{Agashe:2014kda} & -- &  & --   \\
$m_{B_s} (\mev)$   &  $5366.8(2)$ &   \cite{Agashe:2014kda} & -- &  & --   \\
$m_{B_d} (\mev)$   &  $5279.2(2)$ &   \cite{Agashe:2014kda} & -- &  & --   \\
$\Delta m_K (\text{ps}^{-1)}$   &  $0.005292(9)$ &  \cite{Agashe:2014kda} & -- &  & --   \\
$\Delta m_{B_d} (\text{ps}^{-1})$   &  $0.510(3)$ &  \cite{Amhis:2014hma} & -- &  & --   \\
$\Delta m_{B_s} (\text{ps}^{-1})$   &  $17.757(21)$ &  \cite{Amhis:2014hma} & -- &  & --   \\
$\tau_{H}^{s} (\text{ps})$   &  $1.607(10)$ &   \cite{Amhis:2014hma} & -- &  & --   \\
$|V_{us}|$   &  $0.2249(9)$ &  \cite{Lubicz} &  $6$ & \cite{Andreazza:2015bja} & 6 & \cite{Andreazza:2015bja}   \\
$|V_{cb}|\times 10^3$ & $40.9(11)$  &  \cite{Lubicz}   &   $ 4 $  & \cite{Aushev:2010bq,Andreazza:2015bja} &  3  &   \cite{Aushev:2010bq,Andreazza:2015bja}\\
$|V_{ub}|\times 10^3$ & $3.81(40)$  &  \cite{Lubicz}   &   $ 10 $  & \cite{Aushev:2010bq,Andreazza:2015bja} &  8  &   \cite{Aushev:2010bq,Andreazza:2015bja}\\
$\sin 2\beta$ & $0.679(20)$\hfill  &  \cite{Amhis:2014hma}  & $16$ & \cite{Aushev:2010bq,Andreazza:2015bja}  &  $8$ & \cite{Aushev:2010bq,Andreazza:2015bja}\\
$\gamma$  & $(73.2^{+6.3}_{-7.0})^\circ $\hfill &  \cite{Amhis:2014hma} & $3^\circ$ & \cite{Andreazza:2015bja, Aushev:2010bq, Bediaga:2012py} & $1^\circ$ & 
\cite{Andreazza:2015bja, Aushev:2010bq, Bediaga:2012py}\\
${\cal B}(B_s \!\to\!\mu^+\mu^-)\!\times\! 10^9$ & $2.8(7)$ &  \cite{Amhis:2014hma} & $3$ & \cite{Aushev:2010bq,Andreazza:2015bja}  &  $1.3$  & \cite{Aushev:2010bq,Andreazza:2015bja}\\
${\cal B}(K^+ \!\to\! \pi^+\nu\bar\nu) \!\times\! 10^{11} $ &  $17.3^{+11.5}_{-10.5}$  & \cite{Amhis:2014hma} &  $0.8$  & \cite{Aushev:2010bq,Andreazza:2015bja}  &  $0.4$ & \cite{Aushev:2010bq,Andreazza:2015bja}\\
${\cal B}(\klpn) \!\times\! 10^{11} $ &  $-$  & &  $2$  & \cite{Aushev:2010bq,Andreazza:2015bja}  &  $0.3$ & \cite{Aushev:2010bq,Andreazza:2015bja}\\
$|\eps_K| \times 10^{-3}$\hfill  & 2.228(11) & \cite{Agashe:2014kda} & -- &  & --	\\
$f_K (\mev)$  & $156.3(9)$\hfill & \cite{Lubicz}  &  $6$ & \cite{Andreazza:2015bja} & $4$ & \cite{Andreazza:2015bja}\\
$\hat B_K$  & $0.766(10)$\hfill &  \cite{Lubicz}   &  $7$  & \cite{Andreazza:2015bja} & $4$  & \cite{Andreazza:2015bja}\\
$\kappa_\epsilon$  & $0.94(2)$\hfill  &  \cite{Buras:2010pza} & ? &  & ? & \\
$\eta_B$  & $0.55(1)$\hfill  &  \cite{Buras:1990fn} & $0.5$ &\cite{Gorbahn} & 0.2 & \cite{Gorbahn}\\
$\eta_{cc}$ & $1.87(76)$ \hfill &  \cite{Brod:2011ty} & ? &   & ? &\\
$\eta_{ct}$ & $0.496(47)$\hfill  &  \cite{Brod:2010mj} & ? &   & ? & \\
$\eta_{tt}$ & $0.5765(65)$\hfill &  \cite{Buras:1990fn} & $30$ & \cite{Gorbahn} & 10 & \cite{Gorbahn}\\
$\delta P_c(X) / P_c(X)$ & $0.408(24)$  &  \cite{Brod:2008ss,Isidori:2005xm} &  $?$ &  & ? & \\
$f_{B_s} (\mev)$ & $226(5)$ & \cite{Lubicz} &  $2$  & \cite{Andreazza:2015bja} & $1$  & \cite{Andreazza:2015bja}\\
$\hat B_{B_s}$ & $1.33(6)$\hfill  &  \cite{Lubicz} & $2$  & \cite{Andreazza:2015bja} & $0.7$ & \cite{Andreazza:2015bja} \\
$f_{B_s}/f_{B_d}$ & $1.204(16)$  &  \cite{Lubicz}  &  $10$  & \cite{Andreazza:2015bja}  & $5$ & \cite{Andreazza:2015bja}\\
$\hat B_{B_s}/ \hat B_{B_d}$  &  $1.03(8)$ \hfill & \cite{Lubicz} & $2$ & \cite{Andreazza:2015bja} & $0.5$ & \cite{Andreazza:2015bja}
\end{tabular}
\end{center}
\caption{\em Present values and future uncertainties for the most relevant quantities of our analysis. In the predictions for future errors we use the symbol {\rm ``--''} when no significant improvement is expected, and the symbol {\rm ``?''} when improvement is expected but difficult to quantify.}
\label{table:inputs}
\end{table}


\boldmath
\subsection*{$\Delta m_{B_s}$}
\unboldmath

The mass differences of the $B^0_{s,d}$--$\bar B^0_{s,d}$ systems in the SM can be written as~\cite{Buchalla:1995vs}
\be
\label{DMq}
\Delta m_{B_q} = \frac{G^2_F}{6\pi^2} m_{B_q} M^2_W 
\hat B_{B_q} f^2_{B_q} \eta_B S_0(x_t) |V_{tq}V^*_{tb}|^2 \,,\qquad q = d,s\,, 
\ee
where $\eta_B $ accounts for NLO QCD corrections.
The LO loop function $S_0(x_t)$ depends on $x_t = 2y^2_t/g_2^2$, where $g_2$ is 
the coupling of the SM gauge group $SU(2)_L$ and $y_t$ is the top-Yukawa coupling,
and is given by
\be
\label{S0}
S_0(x_t)  = \frac{4x_t - 11 x_t^2 + x_t^3}{4(1-x_t)^2}-\frac{3 x_t^3 \log x_t}{2
(1-x_t)^3}\approx 2.32 
\left(\frac{M_t}{173.34\, \gev}\right)^{1.52}
~.
\ee
The latter equality shows the sensitivity of $\Delta M_{d,s}$ to the top mass in the proximity of its physical value.
From \eq{DMq} we obtain the following value for $\Delta m_{B_s}$ 
\begin{equation}
\label{DMs}
\Delta m_{B_s} =
\frac{16.9 \pm 1.4}{{\rm ps}} \bigg( \frac{\sqrt{\hat B_{B_s}}f_{B_s}}{261\,\mev}\bigg)^2
\left(\frac{M_t}{173.34\,\gev}\right)^{1.52}
\bigg(\frac{|V_{ts}V^*_{tb}|}{0.0401} \bigg)^2 \bigg(\frac{\eta_B}{0.55}\bigg) \,.
\end{equation}
Matching this expression with the measurement of $\Delta m_{B_s}$ reported in table \ref{table:inputs}, we find
\begin{equation}
\label{mt_DMs}
\left(M_t\right)_{\Delta m_{B_s}} = (179.3 \pm 9.7)~{\rm GeV}\,.
\end{equation}
Therefore, the current extraction of $M_t$ from $\Delta m_{B_s}$ is affected by an uncertainty of about $ 5\%$. 

\boldmath
\subsection*{$\Delta m_{B_d}$}
\unboldmath

The SM prediction for $\Delta m_{B_d}$ is
\begin{equation}
\label{DMd}
\Delta m_{B_d} = \frac{0.54\pm 0.08}{{\rm ps}}\bigg( \frac{\sqrt{\hat B_{B_d}}f_{B_d}}{213\,\mev}\bigg)^2 
\left(\frac{M_t}{173.34\,\gev}\right)^{1.52}
\bigg(\frac{|V_{td}V^*_{tb}|}{0.0088} \bigg)^2 
\bigg(\frac{\eta_B}{0.55}\bigg)\,,
\end{equation}
and the corresponding determination of the top mass $M_t$ is
\begin{equation}
\label{mt_DMd}
\left(M_t\right)_{\Delta m_{B_d}} =
(167.0 \pm 16.8)~{\rm GeV}\,,
\end{equation}
with an error at the $9\%$ level. Note that the relevant CKM matrix elements and hadronic parameters entering 
$\Delta m_{B_d}$ are currently less precisely known than those of $\Delta m_{B_s}$ (see table~\ref{table:inputs}) 
and this explains the smaller error on $M_t$ in eq.~(\ref{mt_DMs}) than in eq.~(\ref{mt_DMd}).

\boldmath
\subsection*{$\epsilon_K$}
\unboldmath
The SM prediction for $|\varepsilon_K|$ can be written as \cite{Buchalla:1995vs}
\be
\label{epsilonK}
|\varepsilon_K| \!=\!  -\frac{\kappa_\varepsilon G_{\rm F}^2 f_K^2 m_{K} M_W^2\hat B_K}{6 \sqrt{2} \pi^2 \Delta m_K} ~
{\rm Im}\lambda_t 
\bigg[\, 
{\rm Re}\lambda_t\, \eta_{tt} S_0(x_t) + {\rm Re}\lambda_c \left( \eta_{ct} S_0(x_c,x_t) - \eta_{cc} x_c \right)  
\,\bigg]\,,
\ee
where we have used ${\rm Im}\lambda_t = - {\rm Im}\lambda_c$, see eq.~(\ref{eq:Im_lambda_t}). The multiplicative 
factor $\kappa_\varepsilon$~\cite{Buras:2010pza} arises from long-distance contributions
\footnote{Recently~\cite{Lehner:2015jga} found $\kappa_\epsilon=0.963(14)$ using the most recent lattice QCD inputs.}, 
and the parameters $\eta_{tt}$, $\eta_{ct}$, and $\eta_{cc}$ accounts for QCD corrections. So far, $\eta_{tt}$ has been 
calculated at the NLO while $\eta_{ct}$ and $\eta_{cc}$ at the NNLO~\cite{Brod:2010mj, Brod:2011ty}.
The loop function $S_{0}(x_t)$ is given in eq.~(\ref{S0}) and $S_{0}(x_c,x_t)$ is~\cite{Buchalla:1995vs}
\begin{equation}
\label{S0_ct}
S_{0}(x_c,x_t) = x_c \left[ \log\frac{x_t}{x_c} - \frac{3 x_t}{4(1-x_t)} - \frac{3 x^2_t \log x_t }{4(1-x_t)^2} \right]
\approx 
2.24 \times 10^{-3} \left(\frac{M_t}{173.34\,\gev}\right)^{0.13}
\, ,
\end{equation}
where $x_c=\overline{m}^2_c(m_c)/M^2_W$ and $\overline{m}_c(m_c)$ is the $\overline{\rm MS}$ charm-quark mass.
Inserting the numerical values, we find 
\begin{equation}
\label{epsK}
\frac{|\varepsilon_K|}{10^{-3}} =
(1.56 \pm 0.23) \left(\frac{M_t}{173.34\,\gev}\right)^{1.52} + (0.50 \pm 0.19)
\, ,
\end{equation}
which matches the experimental measurement of $|\varepsilon_K|$ for 
\begin{equation}
\label{mt_epsK}
\left(M_t\right)_{|\varepsilon_K|}
= (185.5 \pm 22.2)~{\rm GeV}\,, 
\end{equation}
with a $12\%$ error.

\boldmath
\subsection*{$B_s\to \mu^+\mu^-$}
\unboldmath

The decay $B_s\to \mu^+\mu^-$ has been observed by a combined analysis of CMS and LHCb data~\cite{combo}. 
Although the experimental error is still quite large, see table \ref{table:inputs}, much progress is expected soon. The SM prediction 
for $\BR(B_s\to \mu^+\mu^-)$ at leading order is~\cite{Buchalla:1995vs}
\begin{equation} 
\BR(B_s\to \mu^+\mu^-) = 
\frac{\alpha_{\rm em}^2(M_Z)\, G_{\rm F}^2 \, m_\mu^2\, f_{B_s}^2\, m_{B_s} \tau_H^s}{16\pi^3\sin^4\theta_{\rm W}}  
\sqrt{1-\frac{m_\mu^2}{m^2_{B_s}}}
 |V_{ts}V^*_{tb}|^2 
Y^2_0(x_t)\,,
\label{Bsmm_th}
\end{equation}
where $\alpha_{\rm em}(M_Z)= 128.952(13)$, see table~\ref{table:inputs}, and $Y_0(x_t)$ is the loop function
\be
\label{S0}
Y_0(x_t)  = 
\frac{x_t}{8} 
\left[      
\frac{x_t - 4}{x_t - 1} + \frac{3 x_t}{(x_t -1)^2}\log x_t
\right]
\approx 0.96
\left(\frac{M_t}{173.34\, \gev}\right)^{1.56}~.
\ee
The NLO QCD corrections have been included in \cite{Buchalla:1998ba}
and found to be very small when using the running $\overline{\rm MS}$ top mass in $Y_0(x_t)$.
The discovery of $B_s\to \mu^+\mu^-$ has motivated improved SM calculations and NNLO 
QCD and NLO electroweak corrections have been computed~\cite{Bobeth:2013uxa}. 
Updating the numerical result of~\cite{Bobeth:2013uxa} by making use of the input parameters 
of table~\ref{table:inputs}, we find
\be
{\BR(B_s\to \mu^+\mu^-)} = 
(3.33 \pm 0.05) \times 10^{-9}\, R_{t\alpha}\, R_s \, ,
\label{bsmm}
\ee
where $R_{t\alpha}$ and $R_{s}$ are
\bea
R_{t\alpha} &=& \left(\frac{\alpha_s(M_Z)}{0.1184}\right)^{-0.18} \left(\frac{M_t}{173.34\,{\rm GeV}}\right)^{3.06}\,,
\\ 
R_s &=& \left( \frac{f_{B_s}}{226\, \mev} \right)^{\! 2}\!
                \left( \frac{|V_{cb}|}{0.0409} \right)^{\! 2}\!
                \left( \frac{|V_{ts}V_{tb}^* /V_{cb}|}{0.980} \right)^{\! 2}
                \frac{\tau_H^s}{1.607\, {\rm ps}}\,, 
\eea
and $|V_{ts}V_{tb}^* /V_{cb}|$ is given in \eq{pirulon}.
Finally, we find 
\be
\label{bsmm_now}
{\BR(B_s\to \mu^+\mu^-)} = 
(3.33 \pm 0.24) \times 10^{-9} \left(\frac{M_t}{173.34\,{\rm GeV}}\right)^{3.06}\,, 
\ee
where the uncertainty comes mostly from $V_{cb}$ and, to a lesser extent, from $f_{B_s}$.
Comparing the experimental result for ${\BR(B_s\to \mu^+\mu^-)}$ quoted in table~\ref{table:inputs}
with eq.~(\ref{bsmm_now}), we end up with the following prediction for $M_t$
\begin{equation}
\label{mt_bsmm}
\left(M_t\right)_{B_s\to \mu\mu} = (163.8 \pm 14.7)~{\rm GeV}\,,
\end{equation}
which suffers from an uncertainty of about $9\%$.
%

\boldmath
\subsection*{$\kpn$}
\unboldmath

The branching ratio for $\kpn$ in the SM can be written as~\cite{Buchalla:1995vs}
\begin{align}
\label{bkpnn}
\BR(K^+\to\pi^+\nu\bar\nu) = \tilde \kappa_+ 
\left[
\left(\frac{{\rm Im}\lambda_t}{\lambda^5}X(x_t)\right)^2 +
\left(\frac{{\rm Re}\lambda_c}{\lambda}P_c(x_c)+\frac{{\rm Re}\lambda_t}{\lambda^5}X(x_t)\right)^2
\right],
\end{align}
where $\tilde\kappa_+$ accounts for the hadronic matrix element, which can be extracted from 
the semi-leptonic decays of $K^+$, $K_L$ and $K_S$ mesons~\cite{Mescia:2007kn}, and electromagnetic corrections 
\begin{equation}\label{kapp}
\tilde\kappa_+ = (5.155\pm 0.025 )\times 10^{-11}\left(\frac{\lambda}{0.2249}\right)^8 \!\! (1+\Delta_\text{EM})\,,
\end{equation}
with $\Delta_{\rm EM} = -0.003$. $X(x_t)$ and $P_c(x_c)$ are the loop functions for the top and charm quark contributions. 
The value of $P_c(x_c)$ is given by
\begin{equation}
P_c(x_c) = P^{\rm SD}_c(x_c) +  \delta P_{c,u}  = 0.408 \pm 0.024\,,
\end{equation}
where $P^{\rm SD}_c(x_c) = 0.368 \pm 0.013$, obtained from the results of~\cite{Brod:2008ss} using the inputs of table \ref{table:inputs}, 
and $\delta P_{c,u} = 0.04 \pm 0.02$~\cite{Isidori:2005xm} arise from short-distance (NNLO QCD and NLO electroweak corrections) and 
long-distance contributions, respectively. On the other hand, the loop function $X(x_t)$ can be written as 
\be
\label{X_t}  
X(x_t) = X_0(x_t) + \frac{\alpha_s}{4\pi} X_1(x_t) + \frac{\alpha}{4\pi} X_{\rm ew}(x_t)\,,
\ee
where $X_0(x_t)$ accounts for the LO result~\cite{Buchalla:1995vs} 
\be
\label{X_0}  
X_0(x_t)= 
\frac{x_t}{8} \left[ \frac{x_t + 2}{x_t - 1} + \frac{3x_t - 6}{(x_t - 1)^2} \log x_t  \right]
\approx 
1.50 \left(\frac{M_t}{173.34\,{\rm GeV}}\right)^{1.15}\,,
\ee
while $X_1(x_t)$ and $X_{\rm ew}(x_t)$ are relative to NLO QCD and electroweak corrections, respectively. 

The full two-loop electroweak corrections to the top-quark contribution $X_t$ has been computed~\cite{Brod:2010hi},
bringing the theoretical uncertainty related to electroweak effects well below $1\%$. 
A very accurate approximation of the full result is captured by the expression~\cite{Brod:2010hi}
\be
\label{X_t}  
X(x_t) = \left[\eta_X - \frac{\alpha_{\rm em}}{4\pi} \left( A  -B ~ C^{\frac{M_t}{173.34\,{\rm GeV}}}  + D~ \frac{M_t}{173.34\,{\rm GeV}}  \right) \right]X_0(x_t)\,,
\ee
where $\eta_X = 0.985$ stems from NLO QCD corrections, while $A\!\simeq\!B\!\simeq\!1.12$, $C\!\simeq\! 1.15$, and $D\!\simeq\! 0.18$
arise from NLO electroweak corrections.

Using the inputs of table~\ref{table:inputs}, we find the following prediction for $\BR(\kpn)$
\begin{eqnarray}
\BR(\kpn) =  (8.42 \pm 0.61) \times 10^{-11}\,.
\end{eqnarray}
Even if $\kpn$ has been already observed, its experimental resolution is so poor (see table~\ref{table:inputs}) that any 
extraction of $M_t$ from $\kpn$ is meaningless at present. For this reason, we postpone the determination of $M_t$ 
to the next section, where we discuss future theoretical and experimental improvements.

\boldmath
\subsection*{$\klpn$}
\unboldmath

The branching ratio for $\klpn$ in the SM is fully dominated by the diagrams with internal top exchanges, 
with the charm contribution well below $1\%$. It can be written as follows~\cite{Buchalla:1995vs} 
\begin{equation}
\label{bklpn}
\BR(\klpn)=\kappa_L
\left[\frac{{\rm Im}\lambda_t}{\lambda^5}X(x_t)\right]^2\,,
\end{equation}
where $\kappa_L$ accounts for the hadronic matrix element and is given by~\cite{Mescia:2007kn} 
\begin{equation}
\label{KL}
\kappa_L=(2.223\pm 0.013)\times 10^{-10}\left(\frac{\lambda}{0.2249}\right)^8\,.
\end{equation}
Due to the absence of the charm contribution in eq.~(\ref{bklpn}), the theoretical uncertainties in $\BR(K_L\to\pi^0\nu\bar\nu)$ 
arise only from the CKM matrix elements. We find that the current prediction for $\BR(K_L\to\pi^0\nu\bar\nu)$ is
\begin{eqnarray}
\BR(\klpn) =  (2.64 \pm 0.41) \times 10^{-11}\,,
\end{eqnarray}
where we have used the inputs of table~\ref{table:inputs}. 

This process has not been observed yet. 
Future prospects for the extraction of $M_t$ from $\klpn$ will be addressed in the next section.

\subsection*{Global fit}

\begin{figure}[t]
\begin{center}
\includegraphics[width=0.7\textwidth]{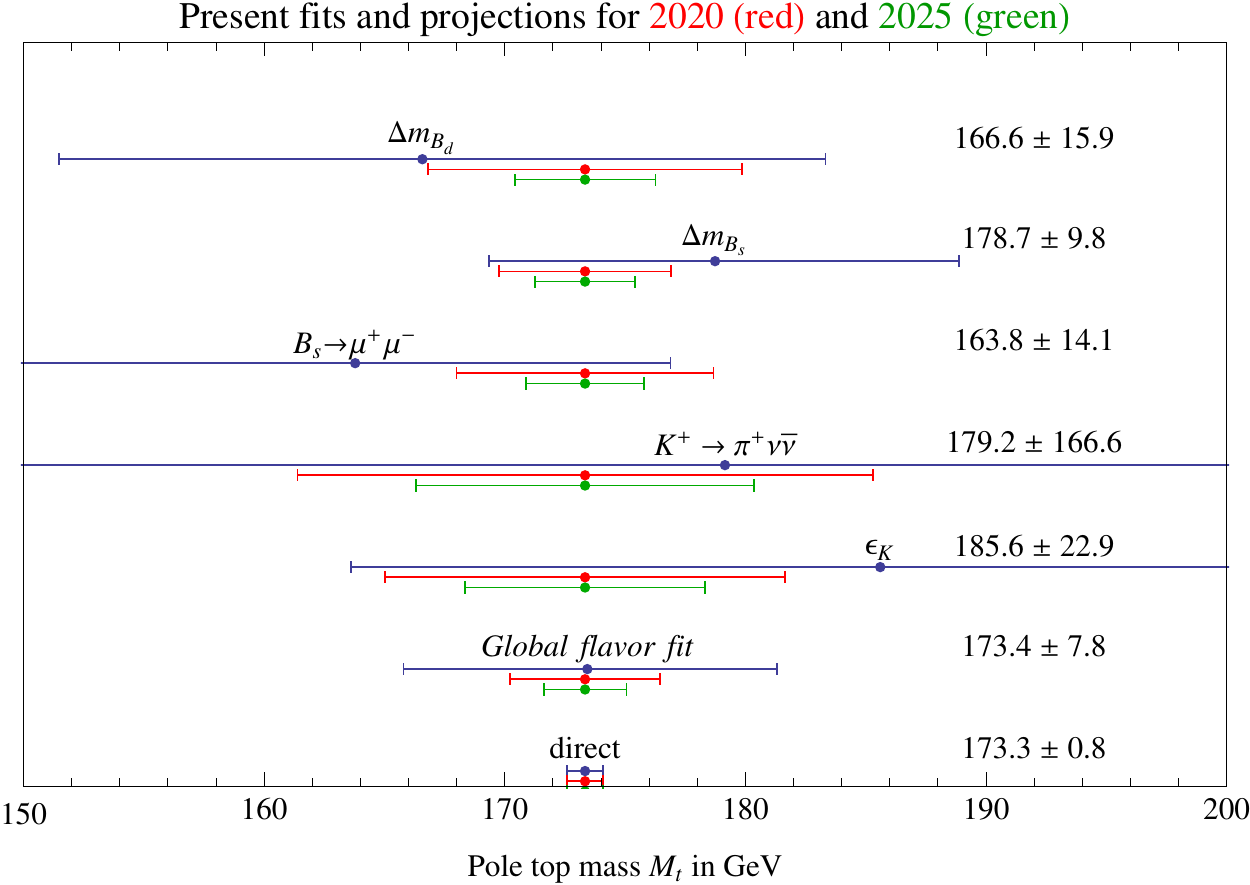}
\caption{\em Summary of present and future determinations of $M_t$ from flavour data. For future projections, we have fixed the central value of $M_t$ to the present direct measurement.
\label{Mtfl}}
\end{center}
\end{figure}

The determinations of $M_t$ from the various flavour processes and their combination are summarised in fig.~ \ref{Mtfl}. Our result for the pole top mass extracted from flavour physics is
\beq\bbox{
(M_t)_{\rm flavour} = (173.4 \pm 7.8)~ \gev } ~ . 
\label{topfl}
\eeq
This result is compatible with the collider determination in \eq{topm}, but the error is too large to be competitive.

In principle, the extraction of $(M_t)_{\rm flavour}$ would require a global fit of all flavour observables in which the CKM parameters and the top mass are allowed to float independently. 
However, in practice, our procedure of fixing the CKM parameters in \eq{ckm_parameters_ours} from processes that are insensitive to $M_t$ and then determine $M_t$ from the remaining 
observables is perfectly adequate and leads to results identical to those from a global fit. Actually, as shown in fig.~\ref{Mtfl}, the determination of $M_t$ is dominated by $\Delta m_{B_s}$, 
which depends on the CKM parameters only through the combination $|V_{ts}V_{tb}^*|$. Equation~(\ref{pirulon}) shows that this combination is equal to $\vcb$, up to a dependence on the 
angles $\gamma$ and $\beta$ suppressed by two powers of $\lambda$. This means that essentially $\vcb$ alone drives the error on the determination of $M_t$ attributable to CKM elements, 
while the less precisely known parameters $\gamma$ and $\beta$ play only a minor role. 
As we will show in the next section, $B_s\to \mu^+\mu^-$ will soon become an equally important process for the determination 
of $M_t$ and its CKM dependence, as in the case of $\Delta m_{B_s}$, is given by $|V_{ts}V_{tb}^*|$. So our conclusion that $\vcb$ is the most important CKM parameter for $M_t$ extraction 
is likely to hold true even after future theoretical and experimental improvements. Let us turn now to discuss our forecast for the future of $M_t$ determinations from flavour processes.

\section{Future determinations of $M_t$ from flavour}\label{Mt_future}

\begin{figure}[t]
\begin{center}
\includegraphics[width=0.45\textwidth]{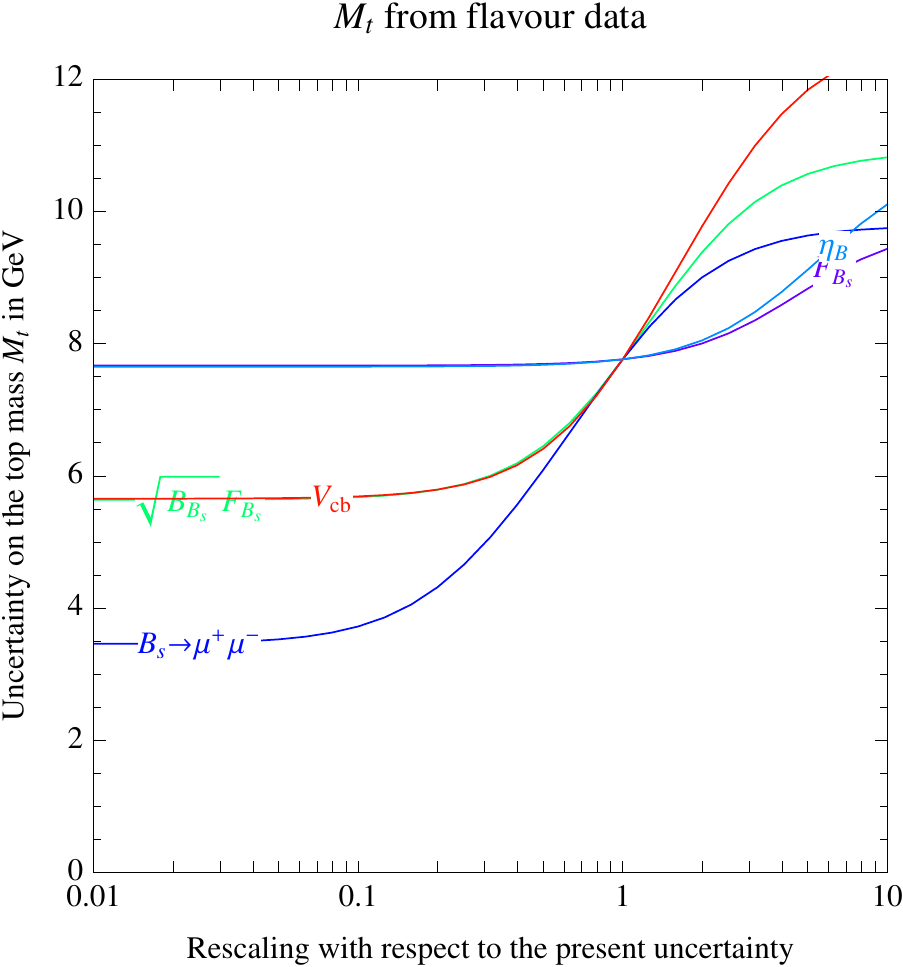}\qquad\includegraphics[width=0.45\textwidth]{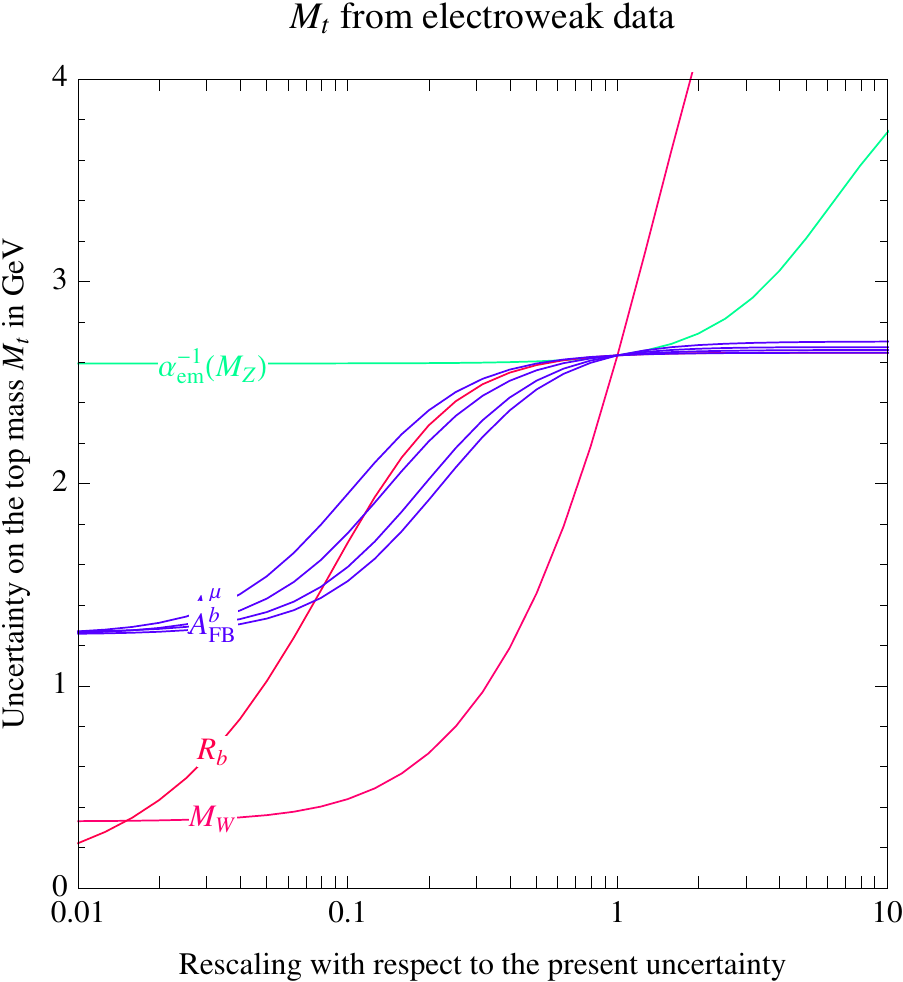}
\caption{\em  Uncertainty on the value of $M_t$ extracted from flavour (left) and electroweak data (right)
as the uncertainties on some key observables are varied one at a time.
\label{Mtfut}}
\end{center}
\end{figure}

The current determination of $M_t$ from flavour processes in eq.~(\ref{topfl}) will soon improve thanks
to upcoming experimental and theoretical progress. 
Figure~\ref{Mtfut} (left panel) shows how the uncertainty on the value of $M_t$ extracted from the global fit changes, 
as we vary the uncertainties of each observable one at a time. We only show the effect of the input parameters
that have a significant impact. We see that more precise measurements of $\BR(B_s\to \mu^+\mu^-)$ and 
$V_{cb}$, and a more precise computation of $\hat B^{1/2}_{B_s} f_{B_s}$ are the key elements for improvements 
in the determination of $M_t$. 

However, future improvements will come simultaneously from many observables.
Thus, in this section we estimate the future situation, in light of new measurements from LHCb, Belle II, and NA62,  
progress in unquenched lattice QCD calculations, as well as improvements in theoretical calculations of QCD and 
electroweak short-distance effects. We will outline the error budget of each flavour observable 
aiming to quantify the improvements needed to bring the error on $M_t$ at the $1\%$ level.

\boldmath
\subsection*{Future determinations of $M_t$ from $\Delta m_{B_s}$}
\unboldmath

In the case of $\Delta m_{B_s}$, the error budget is
\begin{equation}
\label{DMS}
\delta (\Delta m_{B_s}) = 
\left(\pm 1.07_{{\hat B_{B_s}}^{1/2} f_{B_s}} \pm 0.91_{\vcb} \pm 0.31_{\eta_B} \right)\, {\rm ps^{-1}} \,,
\end{equation}
where, hereafter, we always assume the current SM central values.
Eq.~(\ref{DMS}) clearly shows that the major sources of errors arise from (in order of importance): {\it i)} the hadronic parameters 
$\hat B^{1/2}_{B_s}f_{B_s}$, {\it ii)} the CKM matrix elements $\vts\approx\vcb$ , and {\it iii)} short-distance QCD effects encoded 
in the parameter $\eta_B$. On the other hand, the experimental error on $\Delta m_{B_s}$ is $\pm 0.02_{\rm exp}$, 
thus, totally negligible.

The goal of lattice QCD, concerning the calculation of quantities related to  flavour
physics, is to reach a resolution at the $1\%$ level (or even slightly better) by 2025, see table 1.
However, at this level of precision, one should also consider small effects such as isospin
breaking and electromagnetic effects, which are $\mathcal{O}[(m_d - m_u)/{\Lambda_{\rm QCD}}]$ 
and $\mathcal{O}(\alpha)$, respectively, and thus at the $1\%$ level. First lattice studies of isospin 
breaking and electromagnetic effects have been performed in the last years leading to very promising 
results~\cite{Blum:2010ym,Borsanyi:2014jba,Carrasco:2015xwa}. Moreover, lattice calculations of form factors of exclusive 
semileptonic $B$-decays are crucial to extract $\vcb$ and $\vub$. They are extracted from more noisy 
three-point correlators and imply an extrapolation in the transfer momentum, which is computationally intense. 
For the semileptonic decays $B\to D/D^*\ell\nu$, however, one measures on the lattice the difference
of the form factor from unity ({\it i.e.} the SU(3) or heavy-quark symmetric limit), so that the uncertainty 
on the form factor itself turns out to be smaller.

Concerning $\vcb$, there are discrepancies between its inclusive and exclusive extrapolations from 
tree-level decays and Belle II should resolve this problem.\footnote{In our analysis, we use the average 
of inclusive and exclusive determinations of ref.~\cite{Lubicz},
see table~\ref{table:inputs}, which does not include the most recent calculation of $\vcb_{\rm incl}$ of 
ref.~\cite{Alberti:2014yda}.} Overall, exclusive determinations 
are expected to be more precise because they are easier to perform experimentally and also 
because the calculations of the relevant form factors from lattice QCD are less challenging than in 
the inclusive case. Finally, the error associated with $\eta_B$ will be reduced significantly, at least 
by a factor of 3--4 once NNLO QCD and NLO electroweak calculations will be available.

Assuming the expected improvements by about $2025$, see table~\ref{table:inputs}, we have 
\begin{equation}\label{DMS2}
\delta (\Delta m_{B_s})  =
\left(\pm \, 0.17_{{\hat B_{B_s}}^{1/2}f_{B_s}} \pm 0.25_{\vcb} \pm 0.06_{\eta_B}\right)\,{\rm ps^{-1}} \,, 
\end{equation}
which corresponds to about a factor of 4 improvement in the overall error compared to the current error, see eq.~(\ref{DMS}).
We can determine $\delta\! \left(M_t\right)_{\Delta m_{B_q}}$ imposing that the SM prediction for ${\Delta m_{B_s}}$ matches
its experimental measurement. This leads to the relation 
\begin{equation}
\label{mt_DMs_general}
\frac{\delta\! \left(M_t\right)_{\Delta m_{B_q}}}{M_t} = \pm  \, 0.66 ~
\sqrt{
4\!\left(\!\frac{ \delta{\hat B^{1/2}_{B_q} f_{B_q}} }{\hat B^{1/2}_{B_q} f_{B_q}}\!\right)^{\!\!2}
\!\!+4\!\left(\frac{ \delta |V_{tq}|}{|V_{tq}|}\right)^{\!\!2}\!\!+ \left(\frac{ \delta \eta_B }{\eta_B}\right)^{\!\!2} 
+ \left(\frac{ \delta \Delta m_{B_q}}{\Delta m_{B_q}}\right)^{\!\!2} 
}\,,
\end{equation}
where ${\delta \Delta m_{B_q}}$ refers to the experimental uncertainty on ${\Delta m_{B_q}}$. Finally we find
\begin{eqnarray}
\delta\!\left(M_t\right)_{\Delta m_{B_s}} \approx
\left\{ \begin{array}{l}
\pm \,3.6~{\rm GeV}  \qquad {\rm (2020)} \\
\pm \,2.1~{\rm GeV}  \qquad {\rm (2025)}
\end{array}
\right.\,,
\label{eq:bound}
\end{eqnarray}
in good agreement with our numerical results in fig.~\ref{Mtfut}. These values have to be compared with the current uncertainty 
$\delta\!\left(M_t\right)_{\Delta m_{B_s}} = \pm 10~{\rm GeV}$, see eq.~(\ref{mt_DMs}). 

\boldmath
\subsection*{Future determinations of $M_t$ from $\Delta m_{B_d}$}
\unboldmath

In the case of $\Delta m_{B_d}$, the current error budget is
\begin{equation}\label{DMD}
\delta(\Delta m_{B_d})
=
\left(\pm \, 0.056_{{\hat B_{B_d}}^{1/2} f_{B_d}} \pm 0.029_{\vcb} \pm 0.001_{\beta} \pm 0.048_{\gamma} \pm 0.010_{\eta_B} \right)\, \rm ps^{-1}\,,
\end{equation}
while the experimental error $\pm 0.003_{\rm exp}$ is negligible. 
Many considerations done for $\Delta m_{B_s}$ hold here too, the only difference being that the uncertainties
on $\hat B_{B_d} f^2_{B_d}$ and $\vtd^2$ are larger than in the $\Delta m_{B_s}$ case.
Assuming the expected improvements by $2025$, see table~\ref{table:inputs}, we have 
\begin{equation}\label{DMD}
\delta(\Delta m_{B_d})
=
\left(\pm \, 0.008_{{\hat B_{B_d}}^{1/2} f_{B_d}} \pm 0.008_{\vcb} \pm 0.001_{\beta} \pm 0.007_{\gamma} \pm 0.002_{\eta_B} \right)\, \rm ps^{-1}\,,
\end{equation}
which corresponds, as in the $\Delta m_{B_s}$ case, to about a factor of $4$ improvement compared to the current uncertainty.
Notice that now the experimental error $\pm 0.003_{\rm exp}$ is no longer negligible.
The projected errors on $\delta\!\left(M_t\right)_{\Delta m_{B_d}}$ by 2020 and 2025 can be found from eq.~(\ref{mt_DMs_general})
and read
\begin{eqnarray}
\delta\!\left(M_t\right)_{\Delta m_{B_d}} \approx
\left\{ \begin{array}{ll}
\pm \,6.6~{\rm GeV} & {\rm (2020)} \\
\pm \,3.1~{\rm GeV}  & {\rm (2025)}
\end{array}
\right.\,,
\label{eq:bound}
\end{eqnarray}
in good agreement with our numerical results in fig.~\ref{Mtfut}. These values have
to be compared with the current uncertainty $\delta\!\left(M_t\right)_{\Delta m_{B_d}} = \pm 16~{\rm GeV}$,  see eq.~(\ref{mt_DMd}). 
Therefore, by around $2025$, the expected uncertainty on the value of $M_t$ extracted from $\Delta m_{B_d}$ will be about $1.6\%$.

\boldmath
\subsection*{Future determinations of $M_t$ from $B_s\to\mu^+\mu^-$}
\unboldmath

In the case of $\BR(B_s\to\mu^+\mu^-)$, not only theoretical but especially experimental uncertainties 
have to be reduced significantly in order to extract the top mass with an improved accuracy.
On the experimental side, the LHCb collaboration aims at reaching a $10\%$ resolution on 
${\BR(B_s \!\to\! \mu^+\mu^-)}$ in a few years. The final goal, after the LHCb upgrade,
is a resolution around $(4-5)\%$.
On the theoretical side, the main sources of uncertainties arise from the decay constant $f_{B_s}$
and $\vcb$. The error budget for $\BR(B_s\to\mu^+\mu^-)$ is
\be
\frac{\delta\, \BR(B_s\to \mu^+\mu^-)}{10^{-9}} = \pm \,0.05_{\rm th} \pm 0.15_{f_{B_s}} \pm 0.18_{\vcb} \pm 0.02_{\tau^s_H} \,,
\ee
where $\pm \,0.06_{\rm th}$ stems from the estimated error from higher-order effects, as discussed in~\cite{Bobeth:2013uxa}.
On the other hand, the experimental error $\pm \,0.84_{\rm exp}$ is by far dominant at present.
The situation is expected to improve greatly in the future. By 2025 the error budget will be
\be
\frac{\delta\BR(B_s\to \mu^+\mu^-)}{10^{-9}} =  \pm \,0.01_{\rm th} \pm 0.03_{f_{B_s}} \pm 0.05_{\vcb}  \pm 0.02_{\tau^s_H} \,,
\ee
assuming that the errors from higher-order effects will be significantly reduced.

Matching the SM prediction, see eq.~(\ref{bsmm}), with the experimental result leads to
the determination of the top mass uncertainty through the relation
\begin{equation}
\label{mt_Bsmm_general}
\frac{\delta\!\left(M_t\right)_{B_s\to \mu\mu}}{M_t}
= \pm \, 0.33~
\sqrt{
4\!\left(\!\frac{ \delta f_{B_s}}{f_{B_s}}\!\right)^{\!\!2}
\!\!+4\!\left(\frac{ \delta \vcb }{\vcb}\right)^{\!\!2}
\!\!+ \left(\frac{ \delta \tau^s_{H} }{\tau^s_{H}}\right)^{\!\!2}
\!\!+ \left(\frac{ \delta B_{\mu\mu} }{B_{\mu\mu}}\right)^{\!\!2} 
}\,,
\end{equation}
where $\delta B_{\mu\mu}$ stands for the experimental error on $B_{\mu\mu}\equiv \BR(B_s\to \mu^+\mu^-)$.
We predict,
\begin{eqnarray}
\delta\!\left(M_t\right)_{B_s\to \mu\mu}
\approx
\left\{ \begin{array}{ll}
\pm \,5.3~{\rm GeV}  & {\rm (2020)} \\
\pm \,2.4~{\rm GeV}  & {\rm (2025)}
\end{array}
\right.\,,
\label{eq:bound}
\end{eqnarray}
in good agreement with our numerical results in fig.~\ref{Mtfut}.

\boldmath
\subsection*{Future determinations of $M_t$ from $\epsilon_K$}
\unboldmath

In order to reduce significantly the determination from $\epsilon_K$, one would need to improve especially the uncertainties 
on $\eta_{cc}$, $\eta_{ct}$ and $V_{cb}$. It is important to stress that a final answer about the errors on 
$\eta_{cc}$ and $\eta_{ct}$ are expected to come from lattice QCD calculations.
In 2--3 years, a fully controlled calculation reducing the total error coming from $\eta_{cc}$ and $\eta_{ct}$
to the $1\%$ level should be available, although this is a challenging task for lattice simulations.

Let us now study the current error budget of $\epsilon_K$ which is given by
\be
\frac{\delta \epsilon_K}{10^{-3}} = 
\pm \,0.17_{\vcb} \pm 0.14_{\gamma} \pm 0.05_{\beta} \pm 0.04_{{\hat B}^{1/2}_K f_K} \pm 0.15_{\eta_{cc}} \pm 0.08_{\eta_{ct}} \pm 0.02_{\eta_{tt}} \pm 0.04_{\kappa_\varepsilon}
\,.
\ee
On the other hand, the expected error budget by 2025 is
\be
\frac{\delta \epsilon_K}{10^{-3}} = 
\pm \,0.05_{\vcb} \pm 0.02_{\gamma} \pm 0.02_{\beta} \pm 0.02_{{\hat B}^{1/2}_K f_K}  \pm 0.02_{\kappa_\varepsilon,\eta_{cc},\eta_{ct}}\, ,
\ee
where we have assumed that the non-perturbative uncertainties encoded in $\kappa_\varepsilon$, $\eta_{cc}$ and $\eta_{ct}$ will almost disappear thanks 
to lattice calculations~\cite{Christ:2012se}. 
We find the following top mass uncertainties 
\begin{eqnarray}
\delta\!\left(M_t\right)_{\epsilon_K}
\approx
\left\{ \begin{array}{ll}
\pm \,8~{\rm GeV}  & {\rm (2020)} \\
\pm \,5~{\rm GeV}  & {\rm (2025)}
\end{array}
\right.\,,
\label{eq:bound}
\end{eqnarray}
in good agreement with our numerical results in fig.~\ref{Mtfut}.

\boldmath
\subsection*{Future determinations of $M_t$ from $\kpn$}
\unboldmath

In the case of $\BR(K^+ \rightarrow \pi^+ \nu \bar{\nu})$, the by far dominant uncertainty comes
from $\vcb$ and to a lesser extent from the long-distance effects encoded in $P_c(X)$. Concerning the 
latter uncertainty, there is ongoing activity by lattice QCD collaborations aiming to reduce it to the $1\%$ 
level in a few years from now. 

On the experimental side, the NA62 experiment at CERN aims to measure $\BR(\kpn)$ with a $10\%$ accuracy  by 2018 
while a $5\%$ resolution could be the final goal of NA62. The current error budget for  $\BR(\kpn)$ is
\be
\frac{\delta\BR(\kpn)}{10^{-11}} = 
\pm \,0.52_{\vcb} \pm 0.43_{\gamma} \pm 0.02_{\beta} \pm 0.23_{P_c} 
\,.
\ee
By 2025 the error budget will  presumably be
\be
\frac{\delta\BR(\kpn)}{10^{-11}} = 
\pm \,0.14_{\vcb} \pm 0.07_{\gamma} \pm 0.01_{\beta} 
\,,
\ee
where we have assumed that the non-perturbative uncertainties encoded in $P_c$ will disappear thanks to lattice calculations~\cite{Isidori:2005tv}.
We estimate the following future top mass uncertainties 
\begin{eqnarray}
\delta\!\left(M_t\right)_{\kpn}
\approx 
\left\{ \begin{array}{ll}
\pm \,12~{\rm GeV}  & {\rm (2020)} \\
\pm \,7~{\rm GeV}  & {\rm (2025)}
\end{array}
\right.\,, 
\label{eq:bound}
\end{eqnarray}
in good agreement with our numerical results in fig.~\ref{Mtfut}.

\boldmath
\subsection*{Future determinations of $M_t$ from $\klpn$}
\unboldmath

On the experimental side, the KOTO experiment at J-PARC plans to reach the SM level for $\BR(\klpn)$ in a few years from now. 
The expected data  corresponds to a few events. With an upgrade of the KOTO experiment 
the final goal is to obtain a sample of about $100$ SM events, corresponding to a $10\%$ 
resolution on $\BR(\klpn)$.

On the theoretical side, since $\BR(\klpn)$ is fully dominated by short-distance effects, the main sources of uncertainties arise from 
$\vcb$, $\beta$ and $\gamma$, see eq.s~(\ref{eq:Im_lambda_t}),(\ref{bklpn}). In particular, the current error budget for $\BR(\klpn)$ is
\be
\frac{\delta\BR(\klpn)}{10^{-11}} = 
\pm \,0.28_{\vcb} \pm 0.23_{\gamma} \pm 0.19_{\beta} 
\,.
\ee
By 2025 the error budget will be
\be
\frac{\delta\BR(\klpn)}{10^{-11}} = 
\pm \,0.08_{\vcb} \pm 0.04_{\gamma} \pm 0.08_{\beta} 
\,.
\ee
We estimate that the top mass uncertainty will be
\begin{equation}
\label{mt_kpnn_general}
\frac{\delta\!\left(M_t\right)_{\klpn}}{M_t}
= \pm \, 0.43~
\sqrt{
16 \left(\!\frac{ \delta\vcb}{\vcb}\right)^2 \!\!+
4 \left(\!\frac{ \delta R_t\sin\beta}{R_t\sin\beta} \right)^2 \!\!+
\left(\!\frac{ \delta K_{\nu\nu} }{K_{\nu\nu}}\!\right)^2 
}
\,,
\end{equation}
where $\delta K_{\nu\nu}$ is the experimental error on $K_{\nu\nu}\equiv \BR(\klpn)$.
Our projection is
\begin{eqnarray}
\delta\!\left(M_t\right)_{\klpn}
\approx 
\left\{ \begin{array}{ll}
\pm \,57.5~{\rm GeV} &{\rm (2020)} \\
\pm \,9.2~{\rm GeV} & {\rm (2025)}
\end{array}
\right.\,.
\label{eq:bound}
\end{eqnarray}

\begin{figure}[t]
\begin{center}
\includegraphics[width=1.\textwidth]{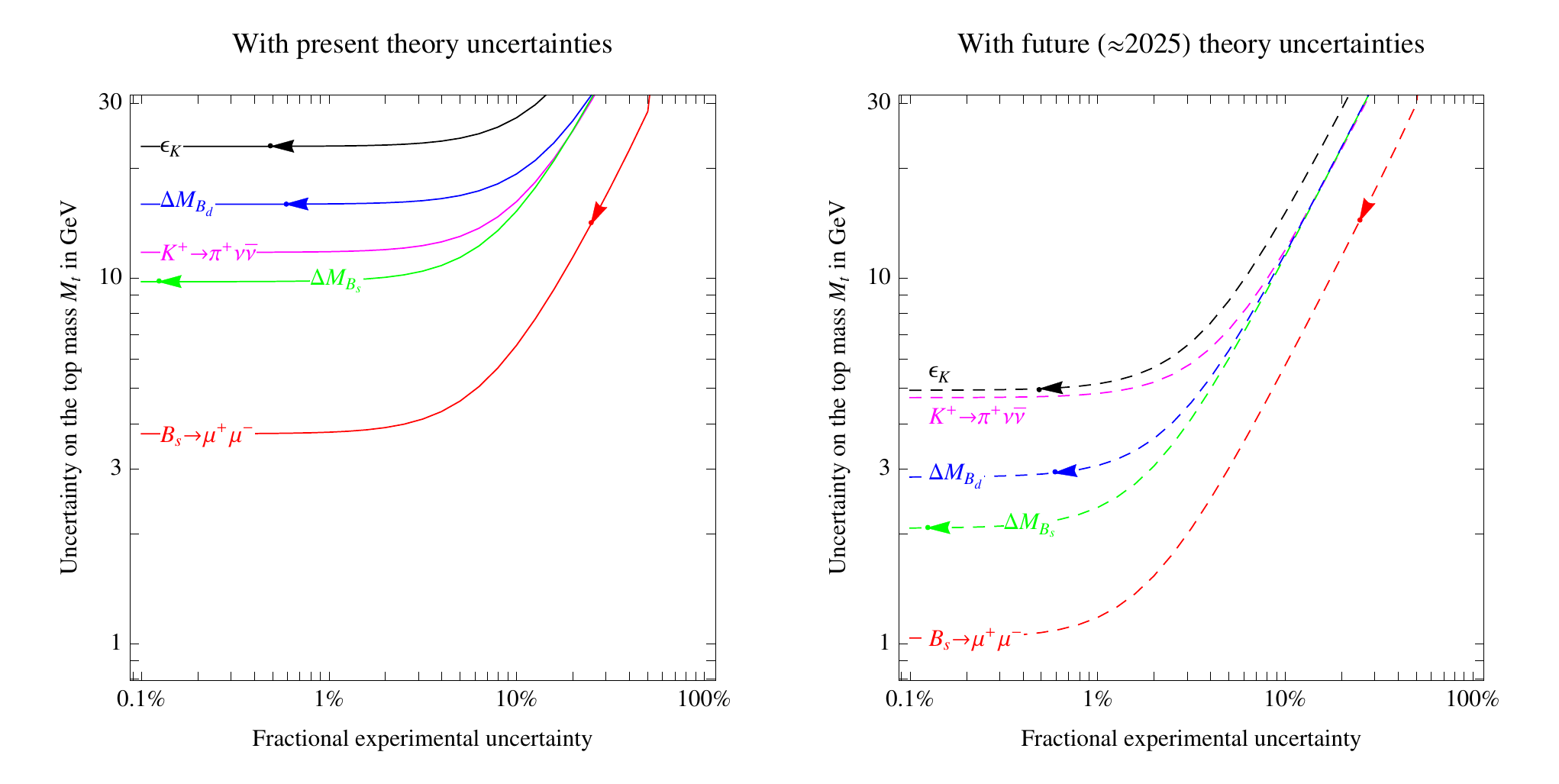}
\caption{\em Uncertainty on the value of $M_t$ extracted from single flavour observables, as the corresponding experimental errors are varied. 
The arrows mark the present experimental uncertainties of the various observables. We assume present theory uncertainties (left) and those 
predicted for around $2025$ (right).}
\label{expfut}
\end{center}
\end{figure}

\boldmath
\subsection*{Global fit}
\unboldmath

The expected determinations of $M_t$ by 2020 and 2025 from the various flavour processes and 
their combination are summarised in fig.~\ref{Mtfl}. From our global fit we predict
\begin{align}
&\delta(M_t)_{\rm flavour} \approx 3.1~ \gev \qquad (2020)\,,
\label{top_2020}\\
&\delta(M_t)_{\rm flavour} \approx 1.7~ \gev  \qquad (2025)\,.
\label{top_2020}
\end{align}
From fig.~\ref{Mtfl} we also learn that $\Delta m_{B_s}$ and  $B_s\to \mu^+\mu^-$ are the most 
accurate $M_t$ discriminators, while other observables like $\Delta m_{B_d}$, $\epsilon_K$ and $K\to \pi \nu\bar\nu$ 
play a sub-leading role.
The latter point is also illustrated by fig.~\ref{expfut}, which shows how experimental improvements in each flavour 
observable affect the uncertainty on $M_t$, assuming present (left) and future (right) theory uncertainties.

\medskip

We are ready now to summarise the main results of this section.

\begin{itemize}

\item Since $ \Delta m_{B_s}$ and $B_s\to \mu^+\mu^-$ are the dominant observables and only depend on the CKM parameters through the combination
$|V_{ts} V_{tb}^*|= \vcb + \mathcal{O}(\lambda^2)$, the determination of $M_t$ essentially does not require a complete global fit analysis.

\item A precise determination of $M_t$ from $\Delta m_{B_s}$ requires substantial improvements 
of $\vcb$ and $\hat B^{1/2}_{B_s} f_{B_s}$, see eq.~(\ref{DMq}).
Concerning $\vcb$, a joint effort of experiments and theory is necessary. 
The measurements of $B\to D/D^*\ell\nu$ branching ratios by Belle II and lattice QCD 
calculations of the relevant form factors should enable us to extract $\vcb$ at or even below the $1\%$ level by around $ 2025$. 
At the same time $\hat B^{1/2}_{B_s} f_{B_s}$ should be calculated by lattice QCD with a precision of about $0.5\%$, see table~\ref{table:inputs}. 
At this level of precision, it will be mandatory to improve also theoretical calculations by the inclusion of NNLO QCD and NLO electroweak short-distance effects. 
On the other hand, the experimental resolution on $\Delta m_{B_s}$, which is already at $0.1\%$, needs not to be improved.
As a result, we expect $\delta\!\left(M_t\right)_{\Delta m_{B_s}} \approx \pm \,2.1~{\rm GeV}$ by about $ 2025$.

\item Unlike the $\Delta m_{B_s}$ case, a precise determination of $M_t$ from $B_s\to\mu^+\mu^-$ requires 
primarily experimental progress in the measurement of its branching ratio. On the theory side, the leading uncertainties
stem from $f_{B_s}$, $\vcb$ and, to a lesser extent, from higher-order effects which are estimated to induce an error
of $1.5\%$~\cite{Bobeth:2013uxa}. By about $2025$, the expected error in $\BR(B_s\to\mu^+\mu^-)$ 
driven by the combination of $f_{B_s}$ and $\vcb$ will be below $2\%$ while the experimental error around 4--5$\%$
and therefore still dominant.  
So, we expect $\delta\!\left(M_t\right)_{B_s\to\mu\mu} \approx \pm \,2.5~{\rm GeV}$ by about $2025$. 
In case the experimental error on $\BR(B_s\to\mu^+\mu^-)$ should reach the $2\%$ level, 
it would be mandatory to improve the estimated $1.5\%$ error associated with higher-order effects. 
In the latter case, we would obtain $\delta\!\left(M_t\right)_{B_s\to\mu\mu} \approx \pm \,1.5~{\rm GeV}$, see fig.~\ref{expfut}.
\end{itemize}
Even though we have identified $\Delta m_{B_s}$ and $\BR(B_s\to\mu^+\mu^-)$ as the dominant $M_t$ discriminators,
we must stress that improvements in the determinations of all other observables discussed in this paper are also important. 
Indeed, our basic assumption for the extraction of the top mass from flavour physics relies on the validity of the SM up to large energy. 
In order to establish whether this situation is realised in Nature or not, we need a global analysis confirming that the CKM picture 
of flavour and CP violation is indeed correct also after the expected theoretical and experimental refinements.

\begin{figure}[t]
\begin{center}
\includegraphics[width=0.6\textwidth]{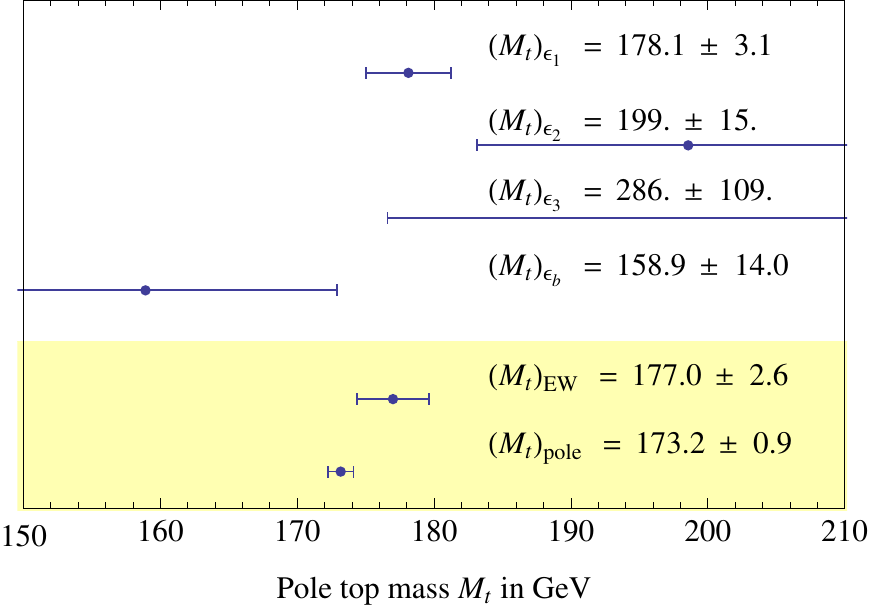}
\caption{\em Summary of present determinations of $M_t$ from electroweak data.}
\label{EW}
\end{center}
\end{figure}

\section{Extracting $M_t$ from electroweak precision data}\label{secEW}

Electroweak observables depend on the top mass (and on the Higgs mass) only through the 
$\varepsilon_1,\varepsilon_2,\varepsilon_3$ parameters that describe corrections to the tree-level propagators of the weak gauge bosons,
and through the $\varepsilon_b$ parameter that describes corrections to the $Zb\bar b$ vertex~\cite{epsilonba}.
These parameters are related to combinations of physical observables and can be extracted from a global fit of experimental measurements:
\begin{equation}
\left\{\begin{array}{l}
\varepsilon_1= +(5.6\pm 1.0)\times 10^{-3}\\
 \varepsilon_{2} = -(7.8\pm 0.9)\times 10^{-3}\\
\varepsilon_{3} =+(5.6\pm 0.9)\times 10^{-3}\\
\varepsilon_b = -(5.8\pm 1.3)\times 10^{-3}
\label{eps123v}
\end{array}\right.
\qquad\hbox{with}\qquad
\rho = \begin{pmatrix}
1 & 0.80 & 0.86 & -0.32\cr 0.80 & 1 & 0.57 & -0.31\cr 0.86&0.57&1 & -0.21 \cr -0.32 & -0.31 & -0.21 &1
\end{pmatrix} \ ,
\end{equation}
where $\rho$ is the correlation matrix.\footnote{The mean values $\mu_i$, the errors $\sigma_i$ and the correlation matrix
$\rho_{ij}$ determine the $\chi^2$ as
$$
\chi^2 =\sum_{i,j} (\varepsilon_i - \mu_i) (\sigma^2)^{-1}_{ij}  (\varepsilon_j - \mu_j),\qquad\hbox{where}
\qquad (\sigma^2)_{ij} = \sigma_i \rho_{ij} \sigma_j\ .$$}
The SM predictions for these observables, for the central values of $\alpha_3(M_Z)$ and $\alpha_{\rm em}(M_Z)$ and
around the measured values of $M_t$ and $M_h$, are\footnote{We thank S.\ Mishima for having provided results of recent computations.}
\begin{equation}
\left\{
\begin{array}{l}
\varepsilon_1= +5.22\times 10^{-3} ~ (M_t/173.34\,{\rm GeV})^{3.15}~~~ (M_h/125.09\,{\rm GeV})^{-0.15}\\
\varepsilon_{2} = -7.32\times 10^{-3}~ (M_t/173.34\,{\rm GeV})^{-0.69}~ (M_h/125.09\,{\rm GeV})^{-0.03}\\
\varepsilon_{3} =+5.28\times 10^{-3}~ (M_t/173.34\,{\rm GeV})^{-0.01}~ (M_h/125.09\,{\rm GeV})^{0.11}\\
\varepsilon_b = -6.95\times 10^{-3}~ (M_t/173.34\,{\rm GeV})^{-2.18}
\label{eps123v}
\end{array}\right. \ .
\end{equation}

As discussed in section~\ref{larget},
in the large $M_t$ limit the one-loop corrections to $\varepsilon_1=\Delta\rho$ and $\varepsilon_b=-2\Delta g^{bb}_{L}$ grow as $M_t^2$,
while $\varepsilon_2$ and $\varepsilon_3$ only have a milder $\ln M_t$ dependence.
Furthermore $\varepsilon_1$ and $\varepsilon_3$ have (in the large $M_h$ limit) a $\ln M_h$ dependence, which leads to a negligible uncertainty, now that 
$M_h = (125.09\pm0.24)\,{\rm GeV}$ is precisely measured. 
Figure~\ref{EW} summarises the various single determinations of $M_t$ from the $\varepsilon$ pseudo-observables, 
from which we derive our result of the global electroweak fit:
\beq \bbox{(M_t)_{\rm EW} =\MtEW} ~ .
\label{pitpit}
\eeq
This result agrees with recent global fits that found $M_t = (177.0\pm2.4)\,{\rm GeV}$~\cite{Baak:2014ora}
and $M_t=(176.6\pm 2.5)\,{\rm GeV}$~\cite{Silvestrini}.
In fig.~\ref{Mtfut} (right panel) we show how the uncertainty on the $M_t$ determination from the global fit changes,
when uncertainties on the various observables are changed one-by-one.  
We only show the effect of those that have the most significant impact.
We see that:
\begin{itemize}

\item  The measurement of $M_W$ plays the key role, since we find $\delta M_t /M_t =69\,  \delta M_W/M_W$. This means that measuring $M_W$ 
with a precision of 8~MeV (as foreseeable after combination of the full LHC dataset~\cite{Baak:2014ora}) can lead to a determination of $M_t$ within 
about 1.2~GeV. On the other hand, measurements of the $WW$ production cross section at the ILC could reduce the error on $M_W$ to about 5~MeV~\cite{Baer:2013cma}, corresponding to a determination of $M_t$ at the level of 0.7~GeV.

\item The determination of $M_t$ can also be improved by better
measurements of the various asymmetries (blue lines in the right panel of fig.~\ref{Mtfut}), which determine the weak mixing angle, and of $R_b$.
However, only with a reduction of the present errors on these quantities by more than a factor of 3 one can start observing meaningful improvements 
on the determination of $M_t$.

\item  The fit is not crucially sensitive to other parameters.  In particular, 
the uncertainty on $M_t$ would be affected only if the error on $\alpha_{\rm em}(M_Z)$ were underestimated by more than a factor of 2.
\end{itemize}

\medskip

Since the determination of $M_t$ from the global fit of electroweak data is largely dominated by $M_W$, it is useful to reconsider the extraction of $M_t$ using $M_W$ as the only input quantity. 
The value of $M_W$ enters the definition of the pseudo-observable $\Delta r_W$, which is defined as the ratio of two different determinations of the weak angle:
 \beq
\Delta r_W \equiv
1 - \frac{\pi\alpha_{\rm em}(M_Z)/\sqrt{2}G_{\rm F}M_Z^2}{M_W^2/M_Z^2(1-M_W^2/M_Z^2)}=(-25.4 \pm 0.95_{M_W} \pm 0.10_{\alpha_{\rm em}})\times 10^{-3} \, .
\label{deltarr}
\eeq
The numerical value has been obtained by taking the experimental values of the SM parameters given in table~\ref{table:inputs}. Equation~(\ref{deltarr}) shows that the uncertainty in $\alpha_{\rm em}(M_Z)$ has a subdominant influence with respect to $M_W$. The quantity $\Delta r_W$ can be computed 
in the SM and expressed in terms of the $\varepsilon$ parameters as
\bea
\Delta r_W &=&   -\tan^{-2}\theta_{\rm W}\, \varepsilon_1 + (\tan^{-2}\theta_{\rm W} -1)\, \varepsilon_2+2\, \varepsilon_3 
\nonumber \\ &=& 
-24.0\times 10^{-3}   \bigg(\frac{M_t}{173.34\,{\rm GeV}}\bigg)^{2.50}\bigg(\frac{M_h}{125.09\,{\rm GeV}}\bigg)^{-0.14}
\,  .
\label{troptrop}
\eea
The above numerical expression has not been obtained by simply replacing \eq{eps123v} into \eq{troptrop}, but rather by using the full two-loop result that can be extracted from the calculation presented in~\cite{Giardino}. Such result has never been included before in global electroweak fits.
By comparing \eq{deltarr} with \eq{troptrop} we find 
\beq (M_t)_{M_W} = (177.7\pm2.8)\,{\rm GeV},\eeq which essentially reproduces the result in \eq{pitpit}, derived from the global fit. This shows that the determination of the top mass from 
electroweak data is almost completely driven by $\Delta r_W$ and a full global fit is superfluous if one is interested in obtaining a simple, but reliable, estimate of $M_t$.

\section{Conclusions}\label{end}
In this paper we have analysed indirect determinations of the top quark mass $M_t$. For this purpose, in section~\ref{larget} we have presented a systematic procedure to identify observables that, 
at the quantum level, have power sensitivity on the top mass, in the limit $M_t\gg M_W$. This is done by considering an effective theory obtained after integrating out the top quark
in the gauge-less limit of the SM. We have divided the physical quantities sensitive to $M_t$ into two classes: flavour observables and electroweak observables.

In section~\ref{flobs} we have discussed how the top mass $M_t$ is determined through $M_t$-dependent quantum effects in the physical quantities
\beq \Delta m_{B_s}\, , \qquad B_s\to \mu^+\mu^-\, ,\qquad
\Delta m_{B_d}\, ,\qquad \epsilon_K\, , \qquad K^+\to \pi^+\nu\bar\nu\, , \qquad \klpn \, .
\label{listobs}
\eeq
The determination of $M_t$ from the first two observables essentially requires only $V_{cb}$ as CKM input. Moreover, these two observables provide the best probe of $M_t$ among flavour processes. Hence, $\Delta m_{B_s}$ and $B_s\to \mu^+\mu^-$, {combined with a determination of $V_{cb}$ and the lattice parameters $\hat B^{1/2}f_{B_s}$ and $f_{B_s}$}, are sufficient to extract a fairly accurate estimate of the $M_t$ determination from flavour physics. Adding to the analysis the other observables listed in (\ref{listobs}) requires a complete joint fit with all CKM parameters and has a limited impact on the extracted value of $M_t$. Our results are summarised in fig.~\ref{Mtfl}:  at present flavour data determine
$M_t = (173.4\pm 7.8)$~GeV, with $\Delta m_{B_s}$ and $B_s\to \mu^+\mu^-$ being the best toppometers.

In section~\ref{Mt_future} we have discussed how the uncertainty on $M_t$ from flavour determinations is expected to decrease significantly in the future,
mostly thanks to better measurements of $B_s\to\mu^+\mu^-$, to better lattice computations of the hadronic parameters entering the SM prediction 
of $\Delta m_{B_s}$ and $B_s\to\mu^+\mu^-$ and to improved theoretical calculations of short-distance effects.
We have estimated that the  uncertainty on $M_t$ can be brought down to $3$~GeV by 2020 and to $1.7$ GeV by 2025.

\medskip

In section~\ref{secEW} we have considered electroweak data, finding that at present they determine $M_t = \MtEW$.  
We have found that $M_W$ and $\Gamma(Z\to b \bar b)$ are the most sensitive quantities, because of the power dependence on $M_t$ of their quantum corrections. 
However, $M_W$ is by far the best toppometer in electroweak physics. We have presented analytic expressions to extract $M_t$ from measurements of $M_W$ which 
take into account recently computed two-loop electroweak quantum corrections~\cite{Giardino}, not yet included in global fit codes.
Figure~\ref{Mtfut} (right panel) shows that a more precise measurement of $M_W$ is the key player for an improved determination of $M_t$ from electroweak observables. 
As experiments at the LHC are expected to reduce the uncertainty on $M_W$ to about 8~MeV~\cite{Baak:2014ora}, it is foreseeable that electroweak physics will 
determine $M_t$ with a precision of about $1.2$~GeV.

In the future, a global fit of all indirect determinations of $M_t$, from both electroweak and flavour data, will provide significant information.
Even if indirect measurements do not surpass direct determinations in precision, the comparison between indirect and direct analyses will carry essential information, 
especially in view of the theoretical ambiguities in the extraction of $M_t$ from collider experiments.

\footnotesize

\subsubsection*{Acknowledgments}
We thank Pier Paolo Giardino for having provided us with the full two-loop prediction for $\Delta r_W$
and Satoshi Mishima for having provided us with recent computations of the $\varepsilon$ parameters.
We thank P. Gambino and T. Mannel for very useful discussions about $V_{cb}$, 
V. Lubicz, G. Martinelli and C. Sachrajda for very helpful discussions about lattice QCD,
and J. Brod, M. Gorbahn, E. Lunghi and M. Misiak for very useful information about short-distance effects.
PP thanks G.  Buchalla, G. Isidori, U. Nierste  and J. Zupan for the invitation
to the MIAPP workshop Flavour 2015: New Physics at High Energy and High Precision,  
where part of his work was performed. The research of PP is supported by the ERC
Advanced Grant No.  267985 (DaMeSyFla), by the research grant TAsP (Theoretical
Astroparticle Physics), and by the Istituto Nazionale di Fisica Nucleare (INFN).


\end{document}